\documentclass[11pt,fleqn,amsmath,amssymb,physrev,pdflatex]{revtex4-2}
\usepackage{amsmath,amssymb} 
\usepackage{bm} 
\usepackage[]{graphicx,color}
\usepackage{grffile} 
\usepackage[]{hyperref} 

\newcommand{\R}{\mathbb{R}} 
 
\newcommand{\C}{\mathbb{C}}

\newcommand{\ii}{\mathrm{i}}

\newcommand{\re}[1]{\mathrm{Re} \left( #1 \right)}
\newcommand{\im}[1]{\mathrm{Im} \left( #1 \right)}
\newcommand{\kc}{K_\mathrm{c}}

\newcommand{\ehopf}{E_\mathrm{hopf}}
\newcommand{\epf}{E_\mathrm{pf}}
\newcommand{\easync}{E_\mathrm{async}}
\newcommand{\edeath}{E_\mathrm{death}}
\newcommand{\dasync}{\delta_\mathrm{async}}
\newcommand{\ddeath}{\delta_\mathrm{death}}
\newcommand{\rl}{R_{\rm lower}}

\renewcommand{\det}[1]{\left|#1\right|}
\DeclareMathOperator{\tr}{Tr}
\graphicspath{{\string~/Works/Projects/CoupledOscillatorsUnderFeedback/img/}}
\graphicspath{{./img/}}

\begin{document}

\title{Feedback-induced desynchronization and oscillation quenching \\in a
population of globally coupled oscillators}

\author{Ayumi Ozawa}
\email{
ozawa-ayumi@g.ecc.u-tokyo.ac.jp
}
\author{Hiroshi Kori}
\affiliation{Department of Complexity Science and Engineering, The University of Tokyo, Chiba 277-8561, Japan}
\date{\today}
\begin{abstract}
   Motivated from a wide range of applications, various methods 
   to control synchronization in coupled oscillators have been proposed. 
   Previous studies 
   have demonstrated that global feedback typically induces 
   three macroscopic behaviors: synchronization, desynchronization, and 
   oscillation quenching.
   However, analyzing all of these transitions 
   within a single theoretical framework is difficult, and thus the feedback effect is 
   only partially understood in each framework.
   Herein, we analyze a model of globally coupled phase oscillators 
   exposed to global feedback, which shows all of the typical 
   macroscopic dynamical states.
   Analytical tractability of the model enables us to obtain 
   detailed phase diagrams where transitions 
   and bistabilities between different macroscopic states are
 identified.
   Additionally, we propose strategies to steer the oscillators into targeted states 
   with minimal feedback strength. Our study provides a useful overview of the effect of global feedback 
   and is expected to serve as a benchmark when more sophisticated
 feedback 
   needs to be designed.
\end{abstract}
\maketitle
\section{Introduction}
Synchronization is a self-organization phenomenon that occurs in interacting oscillatory elements and 
is widely observed\cite{Pikovsky2001book,Glass2001, Arenas2008,winfree01,Kuramoto1984}. 
The resultant coherent oscillation is desirable in some systems.
For example, synchronization is essential for the normal functioning of power grids \cite{Motter2013}. 
Other beneficial effects of synchronization include  
 the enhanced precision in biological oscillatory systems\cite{Enright1980,winfree01,Herzog2004,Needleman2001,Kori2012}, 
coordinated locomotion of animals and robots\cite{Ijspeert2008a}, 
and reduced congestion in models of traffic flow\cite{Lammer2006,Aleko2019}.

However, synchronization may also cause problems. 
At the Millennium Footbridge in London, the steps of the pedestrians were synchronized, 
and considerable lateral movement of the bridge was observed\cite{Strogatz2005}. 
Synchronization is also associated with neurological disorders. 
The local field potentials recorded from the brains of Parkinsonian patients 
and model animals often display marked oscillations, 
which are considered to be reflections of 
coherent neuronal activities\cite{Hammond2007}. 
Although the mechanism of Parkinson's disease is not well understood yet,
exaggerated synchronization is one of the possible factors
that induces related symptoms\cite{McGregor2019}. 
For some types of Parkinsonian patients, deep brain stimulation (DBS), 
which involves electrical stimulation to particular
regions of the brain, 
may suppress the pathological collective oscillation and motor symptoms
\cite{Hammond2007,Armstrong2020}.
However, this treatment is sometimes accompanied by negative side effects\cite{Limousin2019,Bronstein2011}. Thus, milder ways of stimulation need to be developed. 

   The wide range of desirable and undesirable synchronization phenomena 
has drawn substantial research attention to the control of synchronization.
As a control implementation, global feedback loops are known to be effective.
In \cite{Kim2001}, 
the authors experimentally demonstrated that 
a synchronously oscillating state can be stabilized 
in a surface chemical reaction that inherently exhibits 
turbulent oscillatory dynamics.
Global feedback may also desynchronize oscillator assemblies.
Several types of mean-field feedback that efficiently desynchronize oscillators 
have been proposed
\cite{Rosenblum2004, Rosenblum2004a, Popovych2005, Ratas2014,
Tukhlina2007, Luo2009, Franci2012, Popovych2017, Zhou2017a}. 
Moreover, other behaviors  
may be realized through global feedback. 
It is proved that in a particular class of phase oscillators, 
the oscillation death state, in which the mean-field oscillation terminates, 
can be stabilized via global mean-field feedback \cite{Franci2012}. 
Oscillation death is known to appear in various coupled oscillator
systems \cite{Koseska2013}.

These extensive studies \cite{Kim2001,Rosenblum2004, Rosenblum2004a, Popovych2005, Ratas2014, Tukhlina2007, Luo2009,  Popovych2017,Zhou2017a,Franci2012} elucidated that global feedback typically
stabilizes the synchronous, asynchronous, or oscillation-death states.
However, in the oscillator models and feedback forms considered thus far, 
analytically treating all of these macroscopic states is difficult, 
and hence, a comprehensive phase diagram has not been obtained. 
With only partial knowledge being available on the phase diagram, 
feedback control may fail to realize a desired state 
owing to the unexpected stabilization of other states. 
Detailed phase diagrams of an analytically tractable model 
would provide an insight into a general principle of feedback control and 
help design and tune the feedback scheme.

   Herein, we consider the Sakaguchi--Kuramoto model, 
   which describes a population of nonidentical phase oscillators 
   with global coupling, 
   as a coupled-oscillator model 
   and incorporate a global mean-field feedback loop 
   into the system. 
   We show that the system exhibits all three typical macroscopic states: 
   the asynchronously oscillating state, the synchronously oscillating
   state, 
   and the oscillation-death state. 
   By invoking the Ott--Antonsen ansatz\cite{Ott2008,Ott2009}, 
   we comprehensively perform the existence and stability analysis 
   of the macroscopic states, thus obtaining detailed phase diagrams 
   in the space of feedback parameters for different coupling strengths. 
   Our analysis elucidates 
   (i) the dependency of the feedback effect on the parameters 
   of the oscillator model; 
   (ii) the optimal feedback parameters for stabilizing 
   the asynchronous state with minimal feedback strength; and 
   (iii) the existence of the bistability between the oscillation-death state 
   and one of the other two states.
The obtained phase diagrams are numerically validated. 
In addition, we propose a strategy for realizing the asynchronous state 
in the bistable region.
Finally, to support the robustness of our results,
we numerically investigate another model that belongs to a more general class
of oscillator models.

\section{Model} \label{model}
We consider globally coupled phase oscillators under global feedback, given as
\begin{align}
\frac{d\theta_i}{d\tilde t} = \tilde \omega_i + \frac{\tilde K}{N} 
\sum_{j=1}^N \sin\left(\theta_j - \theta_i + \beta\right)
+ \tilde E \sin (\theta_i+\alpha) F(\bm \theta) 
\label{eq:kuramoto-sakaguchi-under-fb-sum-tilde},
\end{align}
where $\theta_i(\tilde t)$ and $\tilde \omega_i$ ($i=1,\ldots,N$) represent 
the phase and natural frequency of the $i$th oscillator, respectively; 
$\tilde K \geq 0$ and $\tilde E \geq 0$ represent the strength of coupling and feedback, respectively; 
and $\alpha$ and $\beta$ are parameters, which are usually nonvanishing
in real oscillator systems\cite{Ashwin2016,Stankovski2017}. 
The function $F(\bm \theta)=F(\theta_1,\ldots,\theta_N)$ describes 
a global feedback and $\sin (\theta_i+\alpha)$ is the phase sensitivity 
to the feedback. In particular, we consider
\begin{align}
 F(\bm \theta) &= \frac{1}{N} \sum_{j=1}^N \cos(\theta_j-\delta),\label{eq:feedback}\\
 &= R \cos(\Theta - \delta),
\end{align}
where $\delta$ is a parameter referred to as the phase offset in the feedback, and 
$R=R(t)$ ($0\leq R \leq 1)$ and $\Theta=\Theta(t)$ $(0\le \Theta < 2\pi)$ are the order parameter and mean phase defined by
\begin{align}
 r := R e^{\ii \Theta} = \frac{1}{N}\sum_{j=1}^N e^{\ii \theta_j}.
\end{align}
The $R$ value indicates the synchronization level.
The complex valued function $r=r(t)$ is referred to as the complex order parameter.
Equation \eqref{eq:kuramoto-sakaguchi-under-fb-sum-tilde}
reduces to the Kuramoto--Sakaguchi model\cite{Sakaguchi1986}
in the absence of feedback, i.e., for $\tilde E=0$. 
See Appendix \ref{sec:derivation} for the derivation of 
Eq.~\eqref{eq:kuramoto-sakaguchi-under-fb-sum-tilde} 
from a general class of coupled limit-cycle oscillators. 
As discussed in Appendix \ref{sec:derivation}, 
 $F(\bm \theta)$ corresponds to a linear 
 function of mean fields in the 
 limit-cycle model 
introduced in Appendix. \ref{sec:derivation}, and 
 the parameters $\tilde{E}$ and $\delta$ can be tuned to arbitrary values 
 when two output signals are observed from individual oscillators.
 Alternatively, it can be implemented 
 when $R(t)$ and $\Theta(t)$ are inferred online.

For analytical tractability, we assume $\tilde \omega_i$ to be drawn from the
Lorentzian distribution $\tilde g\left(\tilde \omega\right)=\frac{\tilde \gamma}{\pi}
\frac{1}{\left(\tilde \omega-\omega_{0}\right)^{2}  +{\tilde \gamma}^{2}}$, where
$\tilde \gamma > 0$ and $\omega_0 > 0$. 
Without loss of generality, we 
decrease the number of parameters 
by introducing nondimensional quantities $t = \omega_0 \tilde t, 
\gamma = \frac{\tilde \gamma}{\omega_0}, K=\frac{\tilde K}{\omega_0}, 
E=\frac{\tilde E}{\omega_0}$, and 
further replacing $\theta_i+\alpha$ by $\theta_i$ for $i=1,\ldots,N$ 
and 
$\delta+\alpha$ by $\delta$.
The resultant equation is 
\begin{align}
\frac{d\theta_i}{d t} = \omega_i
+ \frac{K}{N} \sum_{j=1}^N \sin\left(\theta_j - \theta_i + \beta\right)
+ E \sin\theta_i F(\bm \theta),
\label{eq:kuramoto-sakaguchi-under-fb-sum}
\end{align}
or
\begin{align}
\dot{\theta}_i = \omega_i + K R\sin\left(\Theta - \theta_i + \beta\right) +
 E R \cos(\Theta - \delta) \sin \theta_i,
 \label{eq:kuramoto-sakaguchi-under-fb}
\end{align}
where $\omega_i$ is drawn from
\begin{equation}
 g\left(\omega\right)=\frac{\gamma}{\pi} \frac{1}{\left(\omega-1\right)^{2}  +{\gamma}^{2}}.
\label{eq:lorenzian}
\end{equation}
Now, the mean frequency is set to unity. 
Equation \eqref{eq:kuramoto-sakaguchi-under-fb} with Eqs.~\eqref{eq:feedback} 
and \eqref{eq:lorenzian} is analyzed below. 
There are six parameters involved: 
$N, K>0, E>0, \beta, \delta$, and $\gamma$.

It is known that for $N\to \infty$, a certain class of oscillator assemblies 
including Eq.~\eqref{eq:kuramoto-sakaguchi-under-fb} has a 
low-dimensional manifold, on which a reduced dynamical equation 
can be obtained 
\cite{Ott2008,Ott2009}.
By following \cite{Nagai2010}, we obtain a closed equation for $r$ on
the manifold, given as
 \begin{align}
\dot{r}=\left(-\gamma+\frac{K e^{\ii \beta}}{2}+\ii \right)
 r-\frac{K e^{-\mathrm{i} \beta}}{2}|r|^{2} r-\frac{E R \cos(\Theta - \delta)}{2}\left(1-r^{2}\right). \label{eq:OA}    
\end{align}
Here, we redefined $r= R e^{\ii \Theta}$ as its continuous analogue
\begin{align}
r = \iint \rho\left(\theta,~\omega,~t\right)e^{\ii\theta}d\theta d\omega,
\end{align}
where $\rho\left(\theta,\omega,t\right)d\theta d\omega$ is the fraction of 
the oscillators with natural frequencies 
between $\omega$ and $\omega+d\omega$ and phases 
between $\theta$ and $\theta+d\theta$ at time $t$.
For finite but sufficiently large $N$, 
Eq.~\eqref{eq:OA} is expected to appropriately approximate  
the behavior of $r(t)$ in Eq.~\eqref{eq:kuramoto-sakaguchi-under-fb} after a transient time.

Henceforth, we assume $\left|\beta\right| < \pi/2$; the coupling
promotes synchronization.

\section{Effect of feedback on the macroscopic state}
\label{sec:analytical-result}
\subsection{Classification of macroscopic states} \label{sec:classification}

\begin{figure*}
 \centering
 \includegraphics[width=\textwidth]{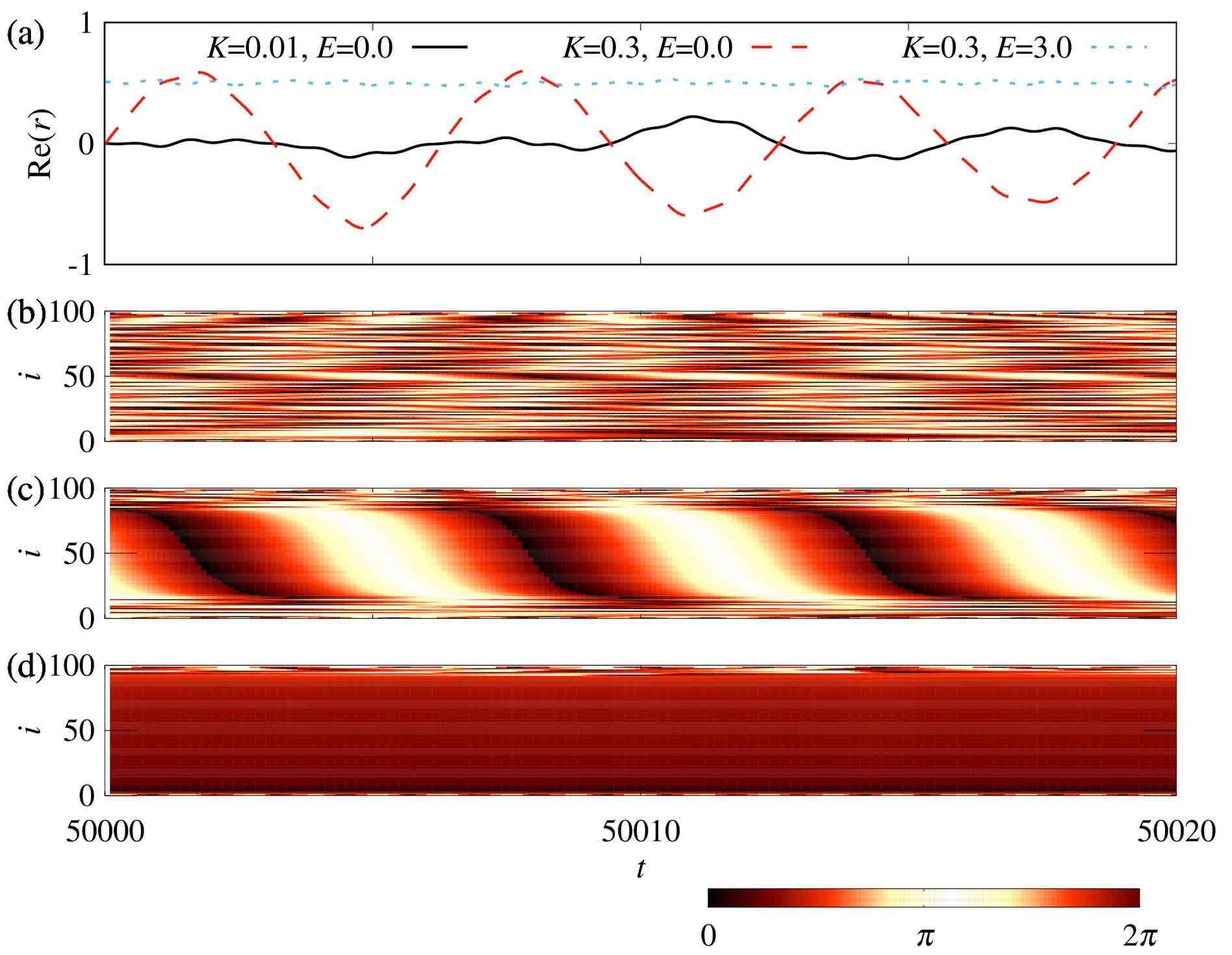}
 \caption{(Color online) Three types of macroscopic state of the
 $N=100$ oscillators described by
 Eq. \eqref{eq:kuramoto-sakaguchi-under-fb}.
(a) Time series of the real part of the complex order
 parameter $r$ for different parameter
 regimes. The black solid, red dashed, and blue dotted
 lines represent the solutions corresponding to the asynchronously oscillating state, 
 the synchronously oscillating state,
 and the oscillation-death state, respectively. 
The coupling strength and feedback parameters are set to $K=0.01$ and
$E=\delta=0$ (solid black line), $K=0.3$ and $E=\delta=0$ (red dashed
 line), and $K=0.3$, $E=3.0$, and $\delta=-\pi/3$ (blue dotted line). The
 other parameters are $\gamma=0.1$ and $\beta=0$.
 (b)--(d) 
 Time series of the individual oscillators. The vertical axis is the index
 of the oscillator, 
and the color scale represents its phase.
The natural frequency $\omega_i$ 
 is set to be monotonically increasing with 
 the oscillator index
 $i$: $\omega_i = 1 + \gamma \tan
 \left[\frac{i\pi}{N} - \frac{\left(N+1\right)\pi}{2N} \right]$.
 The parameters for (b),(c), and (d) are identical to those of the black solid,
red dashed, and blue dotted lines in
 Fig.~\ref{fig:3-macro-states}(a), respectively.
}
 \label{fig:3-macro-states}
\end{figure*}

Dynamical properties of the system described by 
Eq. \eqref{eq:OA} in the absence of feedback ($E=0$) are evident.
This system always has a global attractor for any parameter values
within $K\geq 0$ and $-\frac{\pi}{2} < \beta < \frac{\pi}{2}$. 
There are two global attractors: the steady solution 
   $r(t)=0$ for $K \leq \kc$ and the limit cycle solution 
$r(t)=r_0 e^{i \tilde \omega t}$ for $K > \kc$, where $\kc = 2\gamma/\cos\beta$, $r_0= \left[1-\kc/K\right]^{1/2}$, 
and $\tilde \omega=1+K\sin\beta - \gamma \tan\beta$.
The former corresponds to the asynchronous state, where the oscillators
rotate with their natural frequencies.
The latter corresponds to the synchronously oscillating
state, where a subpopulation of the oscillators is phase-locked to the
oscillating mean field.
Figure \ref{fig:3-macro-states} (a--c) shows typical dynamics of Eq. \eqref{eq:kuramoto-sakaguchi-under-fb}
for $E=0$. 

In addition to these two attractors, a stable steady solution $r(t)=r^* \neq 0$ may arise 
in the presence of the feedback ($E>0$). We refer to the macroscopic state corresponding to 
this solution as the oscillation-death state because of the cessation of 
the microscopic and macroscopic oscillations explained below.
To understand the dynamics of individual oscillators 
in the oscillation-death state ($r(t)=r^*$),
let us consider Eq.~\eqref{eq:kuramoto-sakaguchi-under-fb}.
Inserting $r^* = R^* e^{\ii \Theta^*}$ into Eq. \eqref{eq:kuramoto-sakaguchi-under-fb}, 
we obtain
\begin{align}
 \dot{\theta}_i &= \omega_i + A\sin\left(\theta_i+B\right),
 \label{eq:kuramoto-sakaguchi-under-fb-death}
 \end{align}
where
\begin{align}
 A&={R^*}[\left( E\cos\left(\Theta^*-\delta \right) 
- K\cos\left(\Theta^*+\beta\right)\right)^2 +\left(
K\sin\left(\Theta^*+\beta\right)\right)^2]^{1/2},\\
\tan B &=\frac{ K \sin\left(\Theta^* + \beta\right)}{
E \cos \left( \Theta^* - \delta\right) - K \cos\left(\Theta^* + \beta\right)}.
\end{align} 
A stable fixed point of Eq. \eqref{eq:kuramoto-sakaguchi-under-fb-death}
exists for 
$ \left|\omega_i\right| < \left|A\right|$, 
which implies that the oscillators with 
$\left|\omega_i\right| < \left|A\right|$ 
cease their oscillations.
Thus, the solution $r\left(t\right) =
r^* \neq 0$ corresponds to the oscillation-death state, 
where a subpopulation of the
oscillators and the mean field quit oscillations.
Figure \ref{fig:3-macro-states}(d) and the blue dotted line in
Fig. \ref{fig:3-macro-states}(a) exemplify the dynamics of the
oscillators and the order parameter, respectively, in the oscillation-death state.

\subsection{Bifurcation analysis and phase diagrams} \label{sec:bifurcation}
\begin{figure*}
   \centering
   \begin{minipage}{0.45\textwidth}
   \includegraphics[width=\columnwidth]{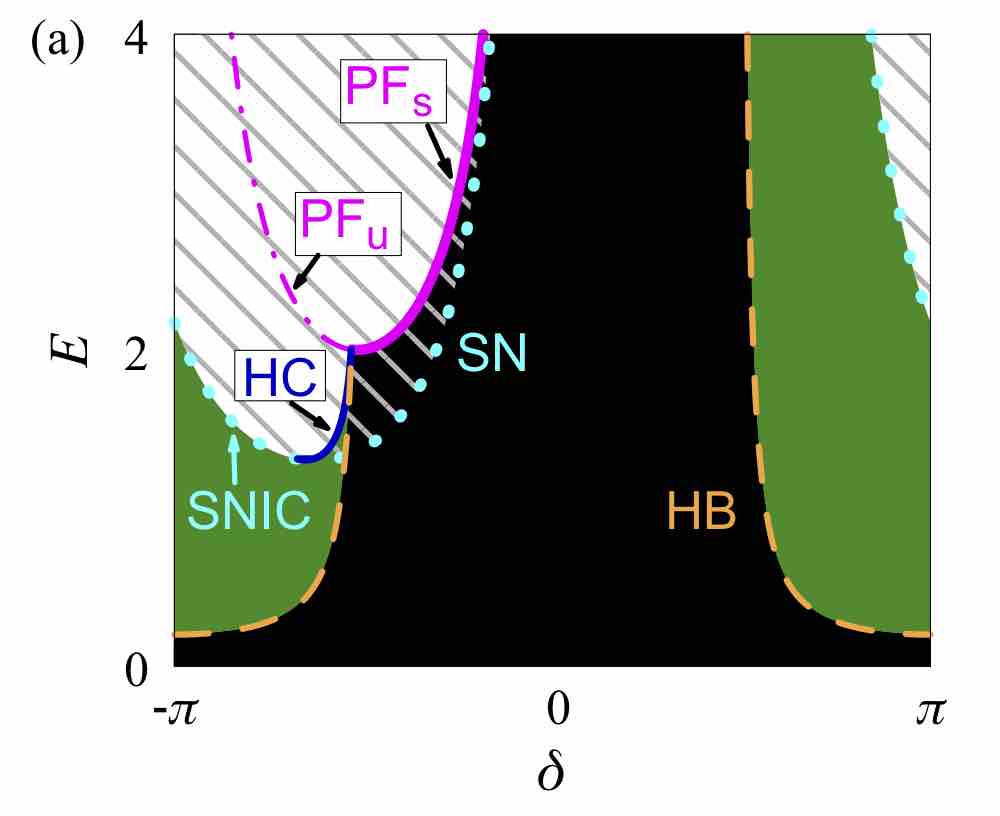}
   \end{minipage}
   \begin{minipage}{0.45\textwidth}
   \includegraphics[width=\columnwidth]{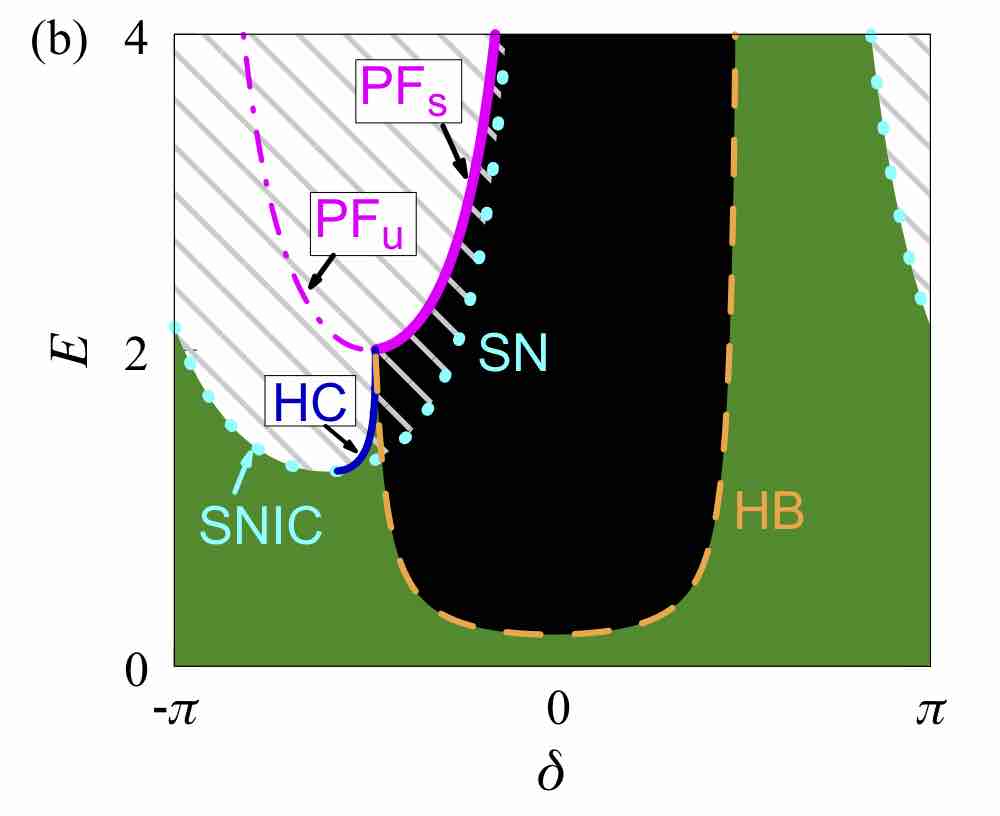}
   \end{minipage}
   \begin{minipage}{0.45\textwidth}
   \includegraphics[width=\columnwidth]{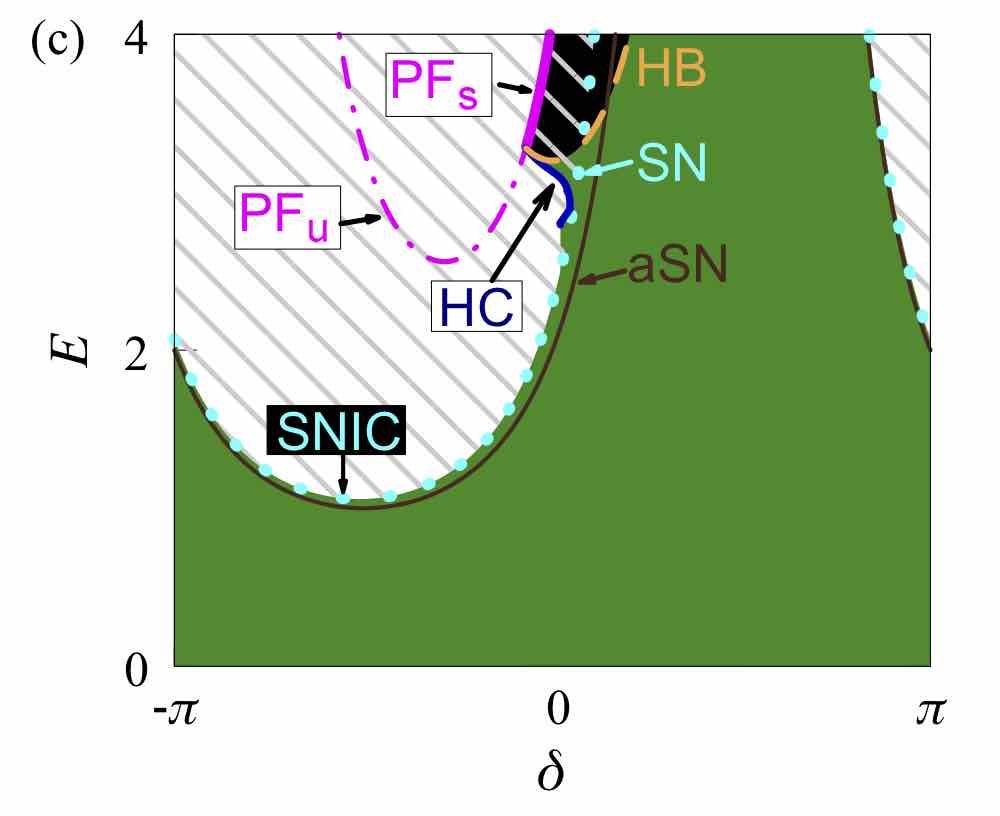}
   \end{minipage}
   \begin{minipage}{0.45\textwidth}
   \includegraphics[width=\columnwidth]{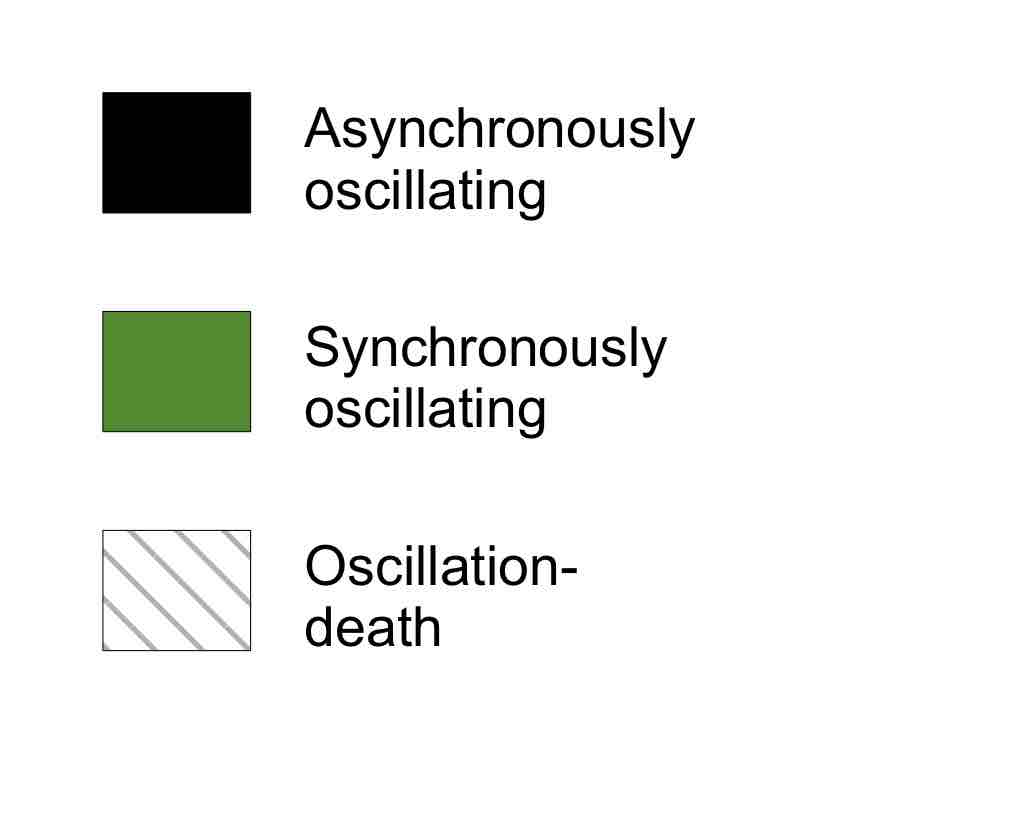}
   \end{minipage}
   \caption{(Color online) Phase diagrams of the macroscopic state based on the stable
solutions of Eq.\eqref{eq:OA} for (a)$K=0.1<\kc$, (b)$K=0.3 > \kc$, and
(c) $K=1.8 \gg \kc$, where $\kc = 2\gamma/\cos\beta = 0.2$. Other
parameters are $\gamma=0.1$ and $\beta=0$. The black, green, and shaded
regions correspond to the asynchronous, synchronously oscillating, and
oscillation-death states, respectively.}
   \label{fig:typical-phase-diagram}
\end{figure*}

The following bifurcation analysis of Eq.~\eqref{eq:OA} enables us to obtain the phase diagrams shown in Fig. \ref{fig:typical-phase-diagram}.

First, we analyze the bifurcation of the fixed point $r=0$, i.e., the asynchronous state.
Substituting $r = x + \ii y~\left(x,~y \in \R\right)$ into
Eq. \eqref{eq:OA} and linearizing the equations for $dx/dt$ and $dy/dt$ around
$\left(x,y\right) = \left(0,0\right)$, 
we obtain the stability matrix 
\begin{align}
 L=\left[
  \begin{array}{cc}
  \Lambda - \frac{E}{2}\cos\delta& -\Omega-\frac{E}{2}\sin\delta\\
   \Omega &\Lambda
  \end{array}
  \right],
\label{eq:stability-matrix-origin}
\end{align}
where $\Lambda=-\gamma+\frac{K}{2}\cos\beta$ and 
$\Omega=1 + \frac{K}{2}\sin\beta$. The quantity $\Omega$ represents the frequency of collective oscillation at its onset, i.e., at $K=\kc$, 
in the absence of feedback.

The saddle-node, transcritical, pitchfork, and Hopf bifurcations are the
codimension-one bifurcations that alter the local stability and/or
number of fixed points.
The first three bifurcations occur when one of the eigenvalues of the stability matrix vanishes,
which can be captured by its necessary condition $\det{L}=0$.
Hopf bifurcation occurs when both eigenvalues of $L$ become purely imaginary,
i.e., $\tr{L}=0$ and $\det{L}>0$.

Solving $\tr{L}=0$ for $E$ under the condition $\det{L}>0$ yields the Hopf
bifurcation curve 
   \begin{align}
      \ehopf= \frac{4\Lambda}{\cos\delta},
      \label{eq:origin-hopf-curve}
     \end{align}
     where $\delta$ satisfies
   \begin{align}
      -\Lambda^2+\Omega^2+2\Lambda\Omega\tan\delta>0.
      \label{eq:origin-hopf-curve-domain}     
   \end{align}
The curve defined by Eqs. \eqref{eq:origin-hopf-curve} and \eqref{eq:origin-hopf-curve-domain} is depicted by the orange dashed
line in Fig.\ref{fig:typical-phase-diagram}(a)--(c). 
Because
$\tr{L} = 2\Lambda\left(1 - E/\ehopf\right)$, 
the asynchronous state is unstable for $E < \ehopf$ (resp. $E>\ehopf$)
when $\Lambda > 0$ (resp. $\Lambda < 0$).
The bifurcation is revealed to be supercritical by the weakly nonlinear analysis performed in 
Appendix \ref{sec:weakly-nlin-analysis}. 
Therefore, a continuous transition between the asynchronous 
and synchronous states occurs on this curve.

The condition $\det{L}=0$ yields another bifurcation curve:
\begin{align}
 \epf=\frac{2\left(\Lambda^2 + \Omega^2\right)}{\Lambda\cos\delta - \Omega\sin\delta}.
 \label{eq:origin-pf-curve}
\end{align}
On this curve, the pitchfork bifurcation occurs 
because Eq.~\eqref{eq:OA} is invariant under the change of $r \to -r$. 
See Appendix \ref{sec:zeroeigen-origin} for the transformation 
 of Eq.~\eqref{eq:OA} to a normal form
   and for 
   a brief explanation
   on why this bifurcation cannot be 
   a transcritical or saddle-node bifurcation.
Because $\det{L}$ can be described as $\det{L} = \left(\Lambda^2 + \Omega^2\right)\left(1 - E/\epf\right)$, 
the asynchronous state is unstable for $E > \epf$ when $\Lambda \Omega
\neq 0$.
The stability of a fixed point changes through this bifurcation 
if the non-zero eigenvalue of the stability matrix is negative; i.e., $\tr L < 0$, or
\begin{align}
2\Lambda < \frac{\Lambda^2+\Omega^2}{\Lambda - \Omega \tan \delta }.\label{eq:origin-pfs-range}
\end{align}
The curve defined by Eqs.~\eqref{eq:origin-pf-curve} and
\eqref{eq:origin-pfs-range}
is higlighted using magenta 
solid line labeled as
PFs in Fig. \ref{fig:typical-phase-diagram}.
On this bifurcation curve, the asynchronous state loses its stability.
    Moreover, weakly nonlinear analysis performed 
    in Appendix \ref{sec:zeroeigen-origin} 
    implies that this bifurcation is subcritical 
    in the parameter region 
    considered in Fig.~\ref{fig:typical-phase-diagram}. 
   This   
    suggests the existence of the bistability between the asynchronous 
    and oscillation-death states near the curve, which will be numerically confirmed 
    in Sec. \ref{sec:num_verification}.

The pitchfork bifurcation occurs to an unstable fixed point 
if the non-zero eigenvalue of the stability matrix on the 
bifurcation curve is positive; i.e., $\tr L > 0$. 
We thus obtain the curve for this bifurcation, which is defined 
by Eq.~\eqref{eq:origin-pf-curve} and Eq.~\eqref{eq:origin-pfs-range} 
but with the opposite inequality. The obtained curve is shown by the dot--dashed 
line labeled as PFu in Fig.~\ref{fig:typical-phase-diagram}. 
Because no stable state emerges with this bifurcation, 
the curve is not relevant to the phase diagram.

Next, we investigate the bifurcation of the fixed point $r^*\neq 0$,
i.e., the oscillation-death state. For convenience, we express
Eq.~\eqref{eq:OA} using polar coordinates: 
\begin{subequations}  
 \label{eq:sk:oa:polar}
  \begin{align}
   \frac{dR}{dt}&=R  \left\{ -\gamma + \left(1-R^2\right)
   \left[\frac{K}{2}\cos\beta- \frac{E}{4}\left(\cos\left(\delta-2\Theta\right)+\cos\delta\right)\right]
   \right\}\label{eq:dot_R} \\
   \frac{d\Theta}{dt}&= 1+\left(1+R ^2\right) \left[\frac{K}{2}  \sin
   \beta+\frac{E}{4}\left(\sin \delta -\sin \left(\delta -2 \Theta \right)\right)\right].
   \label{eq:dot_Theta}  
  \end{align}
\end{subequations}

A fixed point $r^* = R^* e^{\ii \Theta^*}$ 
is given as a solution to
 $dR/dt=0$ and $d\Theta/dt=0$; however, 
this is difficult to solve explicitly.
Nevertheless, we can still obtain the stability matrix by linearizing  Eq.~\eqref{eq:sk:oa:polar} around $(R,\Theta) = (R^*, \Theta^*)$, as
\begin{align}
 M = 
\left[
\begin{array}{cc}
-\frac{2 {R^*}^2 \gamma }{1-{R^*}^2} & -\frac{E}{2}
 {R^*}\left(1-{R^*}^2\right)   \sin (\delta -2 \Theta^* )\\
 -\frac{2 {R^*} }{{R^*}^2+1}& \frac{E}{2} \left({R^*}^2+1\right)
   \cos (\delta -2 \Theta^* )\\
\end{array}
\right].\label{eq:M}
\end{align}
Below, we show that the bifurcation curve can be obtained as a function of $R^*$ 
and drawn in the 
phase diagram 
by varying $R^*$ in the range of $0<R^*<1$.

By solving $\det M = 0$ and using the identity $\sin^2 \left(\delta-2\Theta^*\right)
+ \cos^2\left(\delta-2\Theta^*\right) = 1$, we obtain
\begin{subequations}  
\label{eq:death:sc}
 \begin{align}
  \sin\left(\delta-2\Theta^*\right)=\pm
  \frac{1}{\sqrt{1+\xi^2}},\label{eq:death:s}\\
  \cos\left(\delta-2\Theta^*\right)=\mp \frac{\xi}{\sqrt{1+\xi^2}},\label{eq:death:c}
 \end{align}
\end{subequations}
where 
\begin{align}
   \xi=\frac{ \left({R^*}^2-1\right)^2}{\gamma
   \left({R^*}^2+1\right)^2}. \label{eq:xi}
\end{align}
By substituting Eq.~\eqref{eq:death:sc} into 
Eq. \eqref{eq:sk:oa:polar} and inserting $dR/dt=d\Theta/dt=0$, we obtain
\begin{subequations}  
\label{eq:sn-parametric}
\begin{align}
 E&=\pm\frac{\sqrt{\xi ^2+1} \left((2 \gamma +K {Q_-} \cos \beta)^2+
 \gamma  \xi  (K {Q_+} \sin \beta+2)^2\right)}{\xi  \left(4 \gamma -K
 {Q_-}^2 \cos \beta+\gamma  K {Q_+}^2 \sin \beta\right)},\\
 \cos \delta &= 
 \pm\frac{2 \xi  (2 \gamma +K {Q_-} \cos \beta) \left(4 \gamma -K {Q_-}^2 \cos \beta+\gamma  K {Q_+}^2
   \sin \beta\right)}{\sqrt{\xi ^2+1} {Q_-} \left((2 \gamma +K {Q_-} \cos \beta)^2+\gamma  \xi  (K
   {Q_+} \sin \beta+2)^2\right)}\pm\frac{\xi }{\sqrt{\xi ^2+1}},\\
 \sin \delta &=
 \mp \frac{2 \xi  (K {Q_+} \sin \beta+2) \left(4 \gamma -K {Q_-}^2 \cos \beta+\gamma  K {Q_+}^2 \sin
 \beta\right)}{\sqrt{\xi ^2+1} {Q_+} \left((2 \gamma +K {Q_-} \cos \beta)^2+\gamma  \xi  (K {Q_+}
 \sin \beta+2)^2\right)} \pm \frac{1}{\sqrt{\xi ^2+1}},
\end{align}
\end{subequations}
where $Q_+ =1+{R^*}^2$ and $Q_- = -1 + {R^*}^2$. 
Inserting $0 < R < 1$ into Eq.~\eqref{eq:sn-parametric} yields the cyan
dotted curve in Fig. \ref{fig:typical-phase-diagram}. 
As rationalized below, this curve represents the saddle-node bifurcation
at which stable and unstable oscillation death solutions that exist 
in the area surrounded by the curve collide and disappear.
If transciritical or pitchfork bifurcation occurred on the curve, there should be 
a fixed point $r^* \neq 0$ below the curve. Such a fixed point may not disappear 
unless a bifurcation involving that point occurs, 
thus it should persist 
up to the parameter region of $E=0$. 
However, this contradicts the fact that no 
isolated fixed point except $r=0$ 
exists at $E=0$. Thus, a saddle--node bifurcation occurs on the curve. 
   We now know that a pair of fixed points arise through the
   bifurcation, but their 
   stability remains unclear. Expressing $\tr M$ as a function of $R^*$, we numerically 
   confirmed that $\tr M < 0$ holds on the bifurcation curve.
This implies that a stable node and a saddle collide on the bifurcation curve.
Therefore, we conclude that a stable oscillation-
death state exists inside the curve and disappears on the curve.

It is useful to obtain an approximate expression for the saddle-node bifurcation 
as a function of $\delta$. 
To this end, for some parameter regions, we derive a necessary condition
and a sufficient condition 
for a nonzero fixed point $(R^*, \Theta^*)$ to exist.  
If a nonzero fixed point $(R^*, \Theta^*)$ exists, 
it must satisfy $\left.d\Theta/dt\right|_{\Theta=\Theta^*} = 0$, or, 
\begin{align}
 E= \frac{4 + 2K (1+R^{*2 })\sin\beta}{(1+R^{*2})( \sin (\delta- 2
 \Theta^*) - \sin \delta )}.
\label{eq:death_existence}
\end{align}
The numerator on the right-hand side is positive 
when $\Omega':= 1 + K\sin \beta > 0$. In this case, 
a lower bound $E_{\rm lower} \le E$ 
is given by Eq.~\eqref{eq:death_existence} with $R^{*2}=1$ 
and $\sin (\delta - 2 \Theta^*)=1$, i.e.,
\begin{align}
  E_{\rm lower} = \frac{2 \Omega'}{1-\sin \delta}. \label{eq:SN_lowerbound}
\end{align}
Thus, when $\Omega' > 0$, a necessary condition for 
the existence of a nonzero fixed point is given by 
\begin{align}
   E \ge E_{\rm lower}. \label{eq:death-necessary}
\end{align}

Next, we address a sufficient condition for the existence of a nonzero fixed point.
We restrict ourselves to the parameter region where $\Omega'>0$ and $K > 2\kc$ holds.
Then, for the parameter region under consideration, 
as shown in Appendix \ref{sec:death-sufficient}, 
a nonzero fixed point exists if $E$ satisfies 
both of the following inequalities:
\begin{align}
   E > E_{\rm lower} \left(1 + \eta\right)
    \label{eq:death-sufficient_lower}
\end{align}
and
\begin{align}
   E < 
   \frac{K \cos \beta }{1 + \cos \delta}, \label{eq:death-sufficient_higher}
\end{align}
where 
\begin{align}
   \eta = \frac{\kc/K}{\left(2 - \kc/K\right)\left(1 + K \sin \beta\right)}.
\end{align}
We can find such $E$ for $\beta \simeq 0$ and sufficiently large $K$ 
when a value of $\delta$ is given. See Appendix \ref{sec:death-sufficient} for details.

From Eq.~\eqref{eq:death-necessary} and \eqref{eq:death-sufficient_lower},
for $\Omega' >0$, $\beta \simeq 0$, and sufficiently large $K$, 
it follows that the saddle--node bifurcation  
should occur at $E \in \left[E_{\rm lower}, E_{\rm
lower}\left(1+\eta\right)\right]$.
As $\eta \to 0$ as $K \to \infty$, the bifurcation curve  
is well approximated by $E_{\rm lower}$ for large $K$.  
   In Fig. 2(c), we observe that $E_\mathrm{lower}$, 
   which is depicted as the approximate SN (aSN) curve, 
   approximates the saddle--node bifurcation curve well, 
   although the sufficient condition given by
   Eqs.~\eqref{eq:death-sufficient_lower} and 
   \eqref{eq:death-sufficient_higher} cannot be satisfied in a range of $\delta$.

Similarly to the saddle--node bifurcation curve, 
the Hopf bifurcation curve is obtained 
by imposing $\tr{M} = 0$ and $\det{M}>0$. The former condition yields
\begin{subequations}
   \label{eq:death_hopf}
   \begin{align}
      E&=\sqrt{\frac{\left[Q_-^2 s^2+(c +2 \gamma  (2 {R^*}^2+1))^2
      -4\gamma ^2{R^*}^4\right]^2}{Q_+^2 Q_-^4 s^2}+
      \frac{16 \gamma ^2 {R^*}^4}{Q_+^2 Q_-^2}
      },\label{eq:death_hopf:e}
      \\
      \cos\delta&= \frac{2\left(c+ 2\gamma\left(2 {R^*}^2 + 1\right)\right)}{EQ_+ Q_-},\label{eq:death_hopf:c}\\
      \sin \delta&=-\frac{2 s}{E Q_+}
      -\sqrt{1 - \frac{16 \gamma^2 {R^*}^4}{E^2 Q_+^2 Q_-^2}},\label{eq:death_hopf:s}
   \end{align}   
\end{subequations}
where 
\begin{align}
 &\left(c+2 \gamma  \left(2 {R^*}^2+1\right)\right)^2-4 \gamma ^2 {R^*}^4+s^2 Q_-^2
    \leq 0,\label{eq:death_hopf:nc}\\
s&=K \left(1+{R^*}^2\right) \sin \beta +2,\\
c&=K (-1+{R^*}^4) \cos\beta.
\end{align}
   Inserting $0 < R^* < 1$ into Eqs.~\eqref{eq:death_hopf:e},
   \eqref{eq:death_hopf:c}, and \eqref{eq:death_hopf:s} and requiring
   $\det{M}>0$ and Eq. \eqref{eq:death_hopf:nc} on the curve, 
we obtain the Hopf bifurcation curve. 
As shown in Appendix~\ref{sec:cod2},
 Hopf bifurcation may also  
occur;
however, this is not the case for the parameter region 
considered in Fig. \ref{fig:typical-phase-diagram}.

A codimension-two bifurcation occurs at which the curves PFs and PFu meet.
A normal form of this bifurcation is known \cite{guckenheimer}.
The weakly nonlinear analysis of our system performed in Appendix \ref{sec:cod2}
implies that for $\Omega > 0$, a heteroclinic bifurcation curve extends from this point.
The navy solid curve labeled by HC shows the heteroclinic bifurcation
curve
obtained using the software XPPAUT\cite{Ermentrout2002xpp}.

\begin{figure*}
   \centering
   \includegraphics[width=\textwidth]{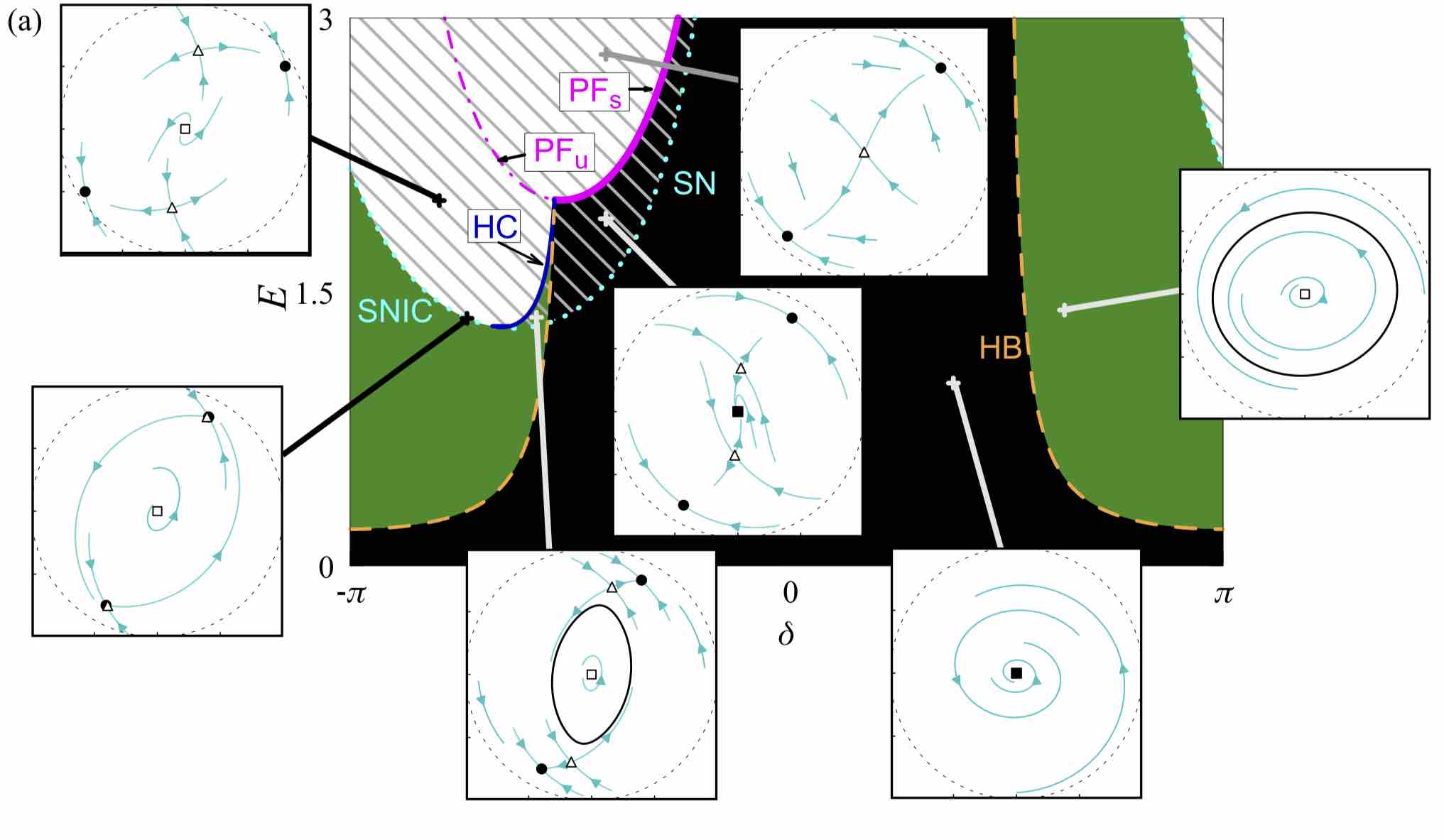}
   \includegraphics[width=\textwidth]{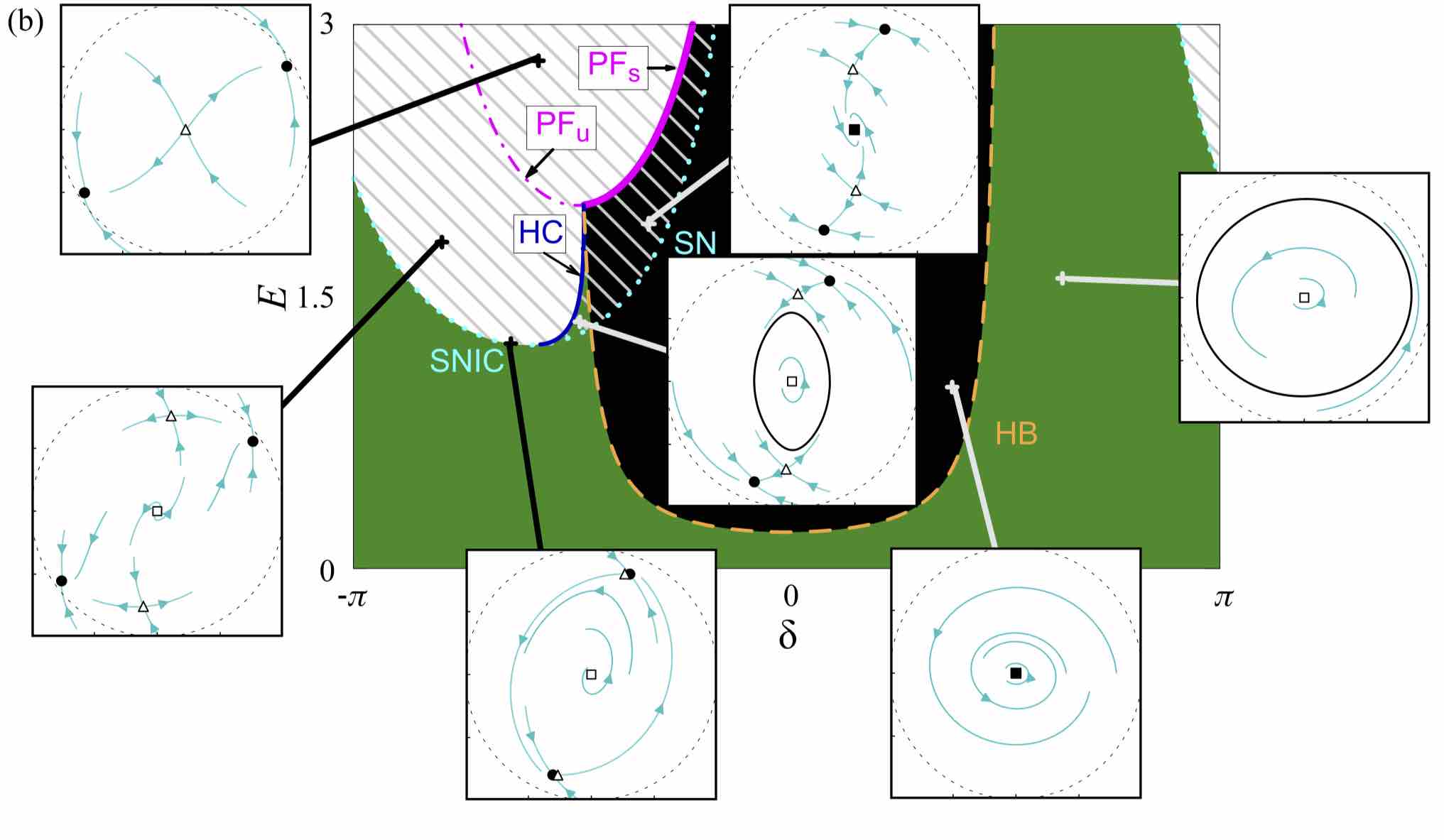}
   \caption{ 
      (Color online) Typical vector fields of Eq.~\eqref{eq:OA} 
    for different areas of the phase diagrams in Fig.~\ref{fig:typical-phase-diagram}.
   The vector fields on the complex planes are drawn with cyan arrows.
   Filled (resp. open) squares represent stable (resp. unstable) spirals, and 
   filled (resp. open) circles represent stable (resp. unstable) nodes.
   The open triangles represent saddles. 
   Stable limit cycles are 
   illustrated using black solid curves.
   The circle drawn with the dashed line on each panel depicts 
   the unit circle on the complex plane. The parameters for (a) and (b) are the same as those shown
   in Fig.~\ref{fig:typical-phase-diagram}(a) and (b), respectively.
 \label{fig:vector-field}
   }
\end{figure*}

Figure \ref{fig:typical-phase-diagram} indicates that the heteroclinic bifurcation curve 
ends up at a point on 
the saddle--node bifurcation curve,
 where another codimension-two bifurcation should occur. 
 The observation of vector fields around this bifurcation point, 
 shown in Fig.~\ref{fig:vector-field}, 
reveals that the saddle--node bifurcation curve is
subdivided into two parts at this point, and the one labeled by SNIC in 
Fig. \ref{fig:typical-phase-diagram} corresponds to saddle--node
bifurcation on an invariant circle. 
A very similar codimension-two bifurcation is reported in \cite{Childs2008}, 
where SN, SNIC, and homoclinic bifurcation curves meet. 
We obtain a heteroclinic bifurcation curve rather than 
a homoclinic one because of the invariance of our system under $r\to -r$. 

The three bifurcation curves HB, HC, and SNIC in Fig.~\ref{fig:typical-phase-diagram} 
form the boundary of the stable synchronous state, 
provided that there is no other bifurcation 
involving periodic solutions, such as the saddle--node bifurcation 
of limit cycles.
Then, the bistable region exists in the area surrounded 
by PFs, HC, and SN. Altogether, 
we obtain three qualitatively 
different phase diagrams 
depending on $K$ values, as shown in Fig. \ref{fig:typical-phase-diagram}.
Our extensive numerical analysis performed in Sec.~\ref{sec:num_verification} and 
Appendix \ref{sec:num-lc} indicates that the phase diagrams given in Fig. \ref{fig:typical-phase-diagram} are comprehensive.

Other bifurcations may occur 
typically for $\Omega<0$
.
As briefly mentioned in Appendix \ref{sec:cod2}, the codimension-two bifurcation 
at the intersection of  
Hopf, PFs, and PFu has a topologically different structure for $\Omega<0$. 
The situation $\Omega < 0$ occurs for $-\frac{\pi}{2} < \beta<0$ and sufficiently large $K$. 
In such a situation, in the absence of feedback, the mean field oscillates 
with a frequency opposite to that of the typical natural frequency of individual oscillators,
 i.e., $\omega_0=1$. This implies that synchronized individual oscillators 
 also have frequencies opposite to their natural ones. 
 However, such a phenomenon is not commonly observed in limit-cycle
 oscillators 
 and should be regarded as an artifact owing to the use of phase approximation 
 for a case of strong coupling. 
 We therefore omit the case of $\Omega < 0$ in the present paper. 

\subsection{Numerical verification } 
\label{sec:num_verification}
\begin{figure*}
   \centering
   \includegraphics[width=17.2cm]{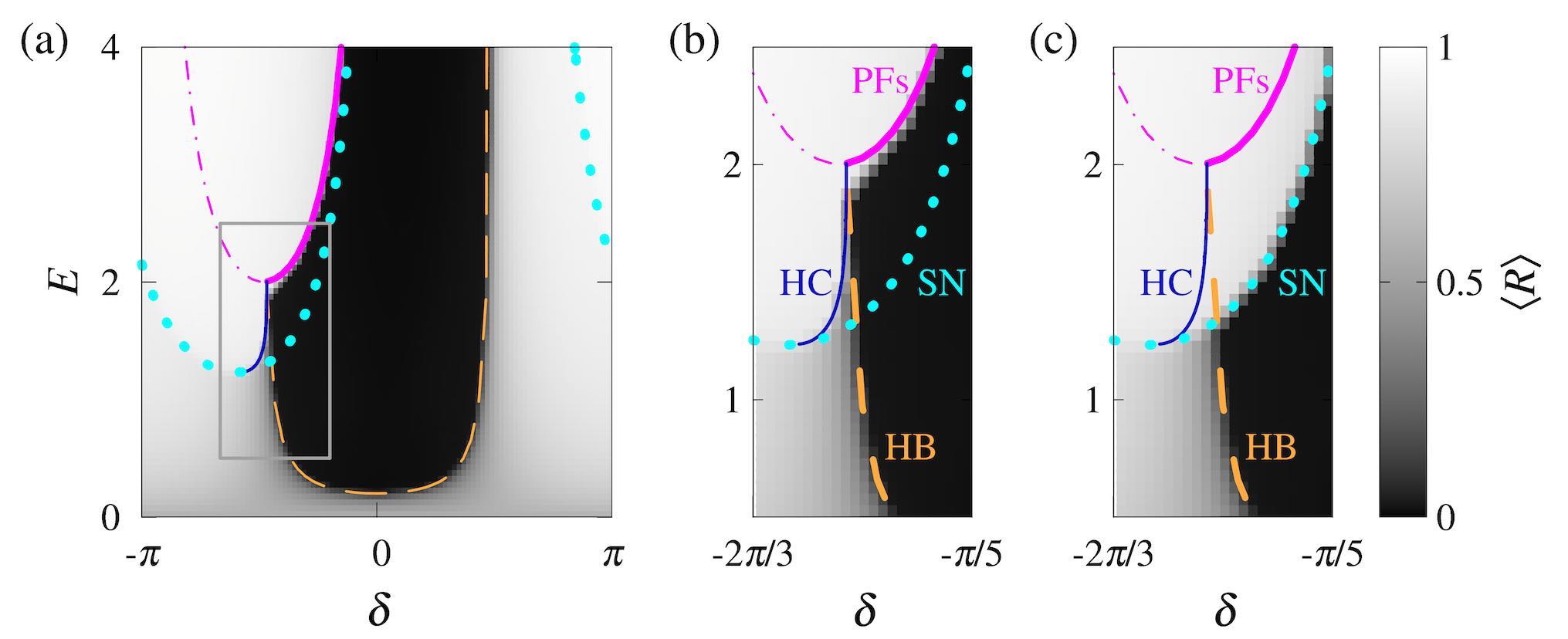}
   \includegraphics[width=17.2cm]{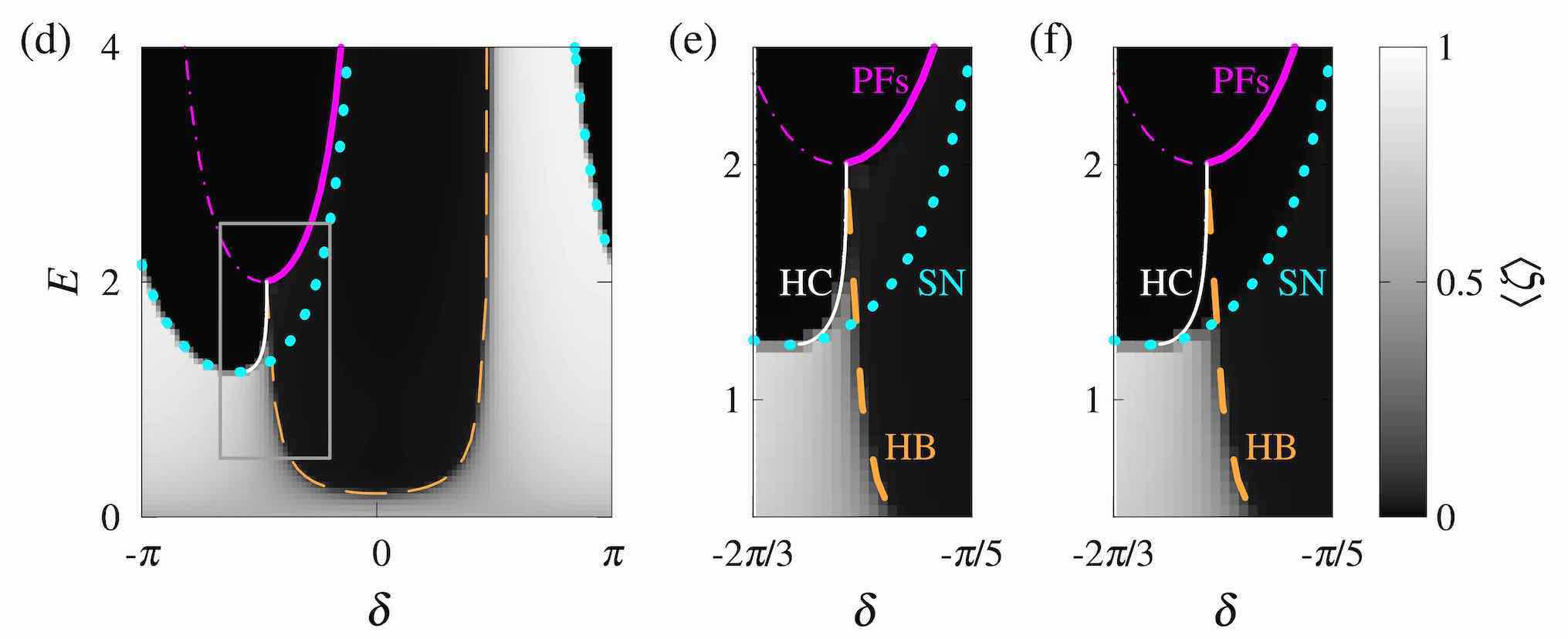}
 \caption{
(Color online) Simulation results of 
Eq.~\eqref{eq:kuramoto-sakaguchi-under-fb}. 
Long-time average of (a--c) $R$ and (d--f) $\zeta$. 
Parameters are the same as those in
 Fig. \ref{fig:typical-phase-diagram}(b), and the same bifurcation curves are drawn here. 
The parameter range in (b,c,e,f) is the same as that in the boxed area in (a,d). 
The initial condition is the uniform phase distribution in (a,b,d,e) 
and the fully synchronized state in (c,f).}
 \label{fig:2000-ks-RandZ}
\end{figure*}

To verify the the analyses in Sec. \ref{sec:bifurcation}, 
we simulated Eq. \eqref{eq:kuramoto-sakaguchi-under-fb} for $N=2000$.
   We set $\omega_i =\omega_0 + \gamma \tan
\left[\frac{i\pi}{N} - \frac{\left(N+1\right)\pi}{2N} \right]$ with $\omega_0=1$, 
which converges to the Lorentzian distribution at $N\to\infty$ \cite{Daido1987}.
In Fig.~\ref{fig:2000-ks-RandZ}(a) and its magnifications 
Fig.~\ref{fig:2000-ks-RandZ}(b) and (c), we show the value of 
$\langle R \rangle$ for different feedback parameters, where the angle brackets denote the long-time average.
As the initial condition, we employ the uniform state, 
i.e., $\theta_i\left(0\right) = 2\pi (i-1)/N$ for $i=1,\cdots,N$, 
in Fig.~\ref{fig:2000-ks-RandZ}(a,b) and 
the fully synchronized state, i.e., $\theta_i\left(0\right)=0$ for $i=1,\cdots,N$, 
in Fig.~\ref{fig:2000-ks-RandZ}(c).
The parameters are same as in Fig. \ref{fig:typical-phase-diagram}(b), and
we draw the same bifurcation curves in 
Fig.~\ref{fig:2000-ks-RandZ}(a--c) for comparison.

In the black regions in Fig.~\ref{fig:2000-ks-RandZ}, 
$\langle R \rangle \simeq 0$ is 
obtained, which indicates that 
the system is in the asynchronous state.
Because the initial condition employed in Fig.~\ref{fig:2000-ks-RandZ}(a,b) 
is considered to be very close to the asynchronous state,
    the asynchronous state should be locally stable in the black region in 
    Fig.~\ref{fig:2000-ks-RandZ}(a,b),
 which is in excellent agreement with our prediction 
 in Fig.~\ref{fig:typical-phase-diagram}(b).
 
Moreover, we can observe a discrepancy between 
Figs.~\ref{fig:2000-ks-RandZ}(b) and (c)
in the region surrounded by the curves SN, HB, and PFs, 
where the bistability between the asynchronous and oscillation-death states is predicted. 
To clarify which region of nonvanishing $\langle R \rangle$ in Fig. \ref{fig:typical-phase-diagram} 
corresponds to the 
 synchronously oscillating or oscillation-death states,
 we further measure $\langle \zeta \rangle$, 
 where $\zeta =  \left| r - \langle r \rangle \right|$. 
From the definition of $R$ and $\zeta$, $\langle R \rangle > 0$
and $\langle \zeta \rangle = 0$ imply that the system is in the
oscillation-death state, 
thus we confirm the predicted bistability as well as the existence 
of the stable oscillation-death state inside the SN and SNIC curves.

\section{Optimal feedback parameters}
We consider $K > \kc$, or equivalently $\Lambda := -\gamma + \frac{K}{2} \cos \beta > 0$, 
for which the system falls into the 
synchronously oscillating state 
in the absence of the feedback, 
and determine the value of the phase offset $\delta$ that minimizes the
required feedback strength $E$ 
to suppress the synchronized oscillations.
We can achieve this by leading the system to 
(i) the asynchronous state and 
(ii) the oscillation-death state. 
Their optimal parameter sets $(\delta, E)$ are denoted by 
(i) $(\dasync, \easync)$ and (ii) $(\ddeath, \edeath)$.

The point $(\dasync, \easync)$ can be determined analytically.
Because the asynchronous state is stable for $E$ above the HB curve and below the PFs curve,
$(\dasync, \easync)$ is given by the minimum of the HB curve.
Further, $\delta=0$ provides the minimum 
when 
Eq.~\eqref{eq:origin-hopf-curve-domain} holds for $\delta=0$, resulting in
\begin{subequations}  
\begin{align}
 \dasync &= 0,\\
 \easync &= 4\Lambda =  -4\gamma + 2K \cos \beta. \label{eq:easync}
\end{align}
\end{subequations}
This is the case when $\Lambda \leq \left|\Omega\right|=|1+\frac{K}{2}\sin \beta|$,
which typically arises for 
small $K$ or large $\tan \beta$.
For $\Lambda > \left|\Omega\right|$, 
the smallest $|\delta|$ 
that satisfies Eq.~\eqref{eq:origin-hopf-curve-domain}
provides the minimum; i.e.,
\begin{subequations}  
\begin{align}
 \dasync &=  \frac{\Omega}{\left|\Omega\right|}\arcsin
  \left(\frac{\Lambda^2-\Omega^2} {\Lambda^2+\Omega^2}\right),\\
 \easync &= \frac{2\left(\Lambda^2+\Omega^2\right)}{\left|\Omega\right|}. \label{eq:easync2}
\end{align}
\end{subequations}

Although $(\ddeath, \edeath)$ can only be numerically determined
using Eqs. \eqref{eq:sn-parametric} and \eqref{eq:death_hopf},
an approximate expression can be obtained from 
\eqref{eq:SN_lowerbound}:
\begin{subequations}  
\begin{align}
 \ddeath &\approx -\frac{\pi}{2},\\
 \edeath &\approx 1 + K \sin\beta. \label{eq:edeath_approx}
\end{align}
\end{subequations}

\begin{figure*}
    \centering
 \includegraphics[width=\textwidth]{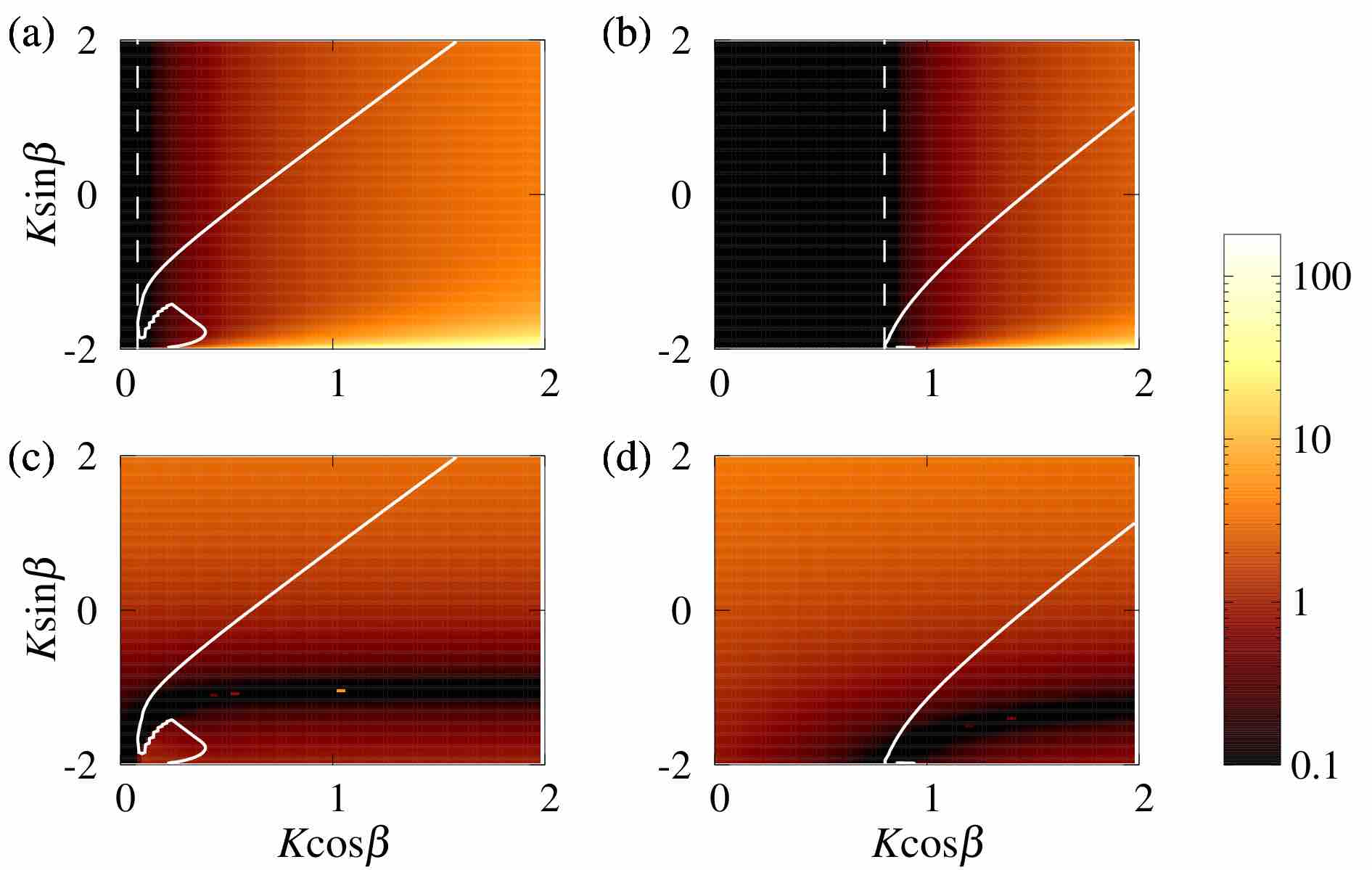}
 \caption{(Color online) Parameter dependency of (a,b) $\easync$ and (c,d) $\edeath$.
 The solid lines represents the parameter values at which $\edeath=\easync$ holds. 
 (a,c) $\gamma=0.04$. (b,d) $\gamma=0.4$. 
 On the left side of the dashed lines in (a) and (b), which depict
 $K\cos\beta=2\gamma$, 
 we have $\easync=0$ because  
 the solution $r=0$ is stable even without feedback.
}
\label{fig:mfs-dependency-on-params}
\end{figure*}

Figure \ref{fig:mfs-dependency-on-params} shows the parameter-dependency of $\easync$ 
given by Eqs.~\eqref{eq:easync} and \eqref{eq:easync2} and $\edeath$ obtained numerically 
using Eqs. \eqref{eq:sn-parametric} and \eqref{eq:death_hopf}. 
The general tendency is well captured by 
Eqs.~\eqref{eq:easync} and \eqref{eq:edeath_approx}.
The solid lines represent the parameter set at which $\easync=\edeath$. 
Based on Eqs.~\eqref{eq:easync} and \eqref{eq:edeath_approx}, 
we can roughly estimate that the asynchronous (oscillation-death) 
state can be achieved with a smaller feedback strength 
when $-4\gamma + 2K\cos \beta$ is small (large)
compared to $1 + K\sin \beta$.

When we desire to suppress the collective oscillation without causing
oscillation death, 
we need to consider the bistability between the asynchronous and oscillation-death states.
For example, see Fig. \ref{fig:typical-phase-diagram}(c), where $\dasync=0$. Suppose that we increase $E$ while fixing $\delta=0$. Then, the oscillation-death state will be obtained before the asynchronous state. By further increasing $E$, we will eventually arrive at the HB curve, above which the asynchronous state is stable. However, because of the bistability, the oscillation-death state is expected to be sustained even in that region. Therefore, to realize the asynchronous state, we need to use a larger $\delta$ value at which we first arrive at the monostable region of the asynchronous state. Once the asynchronous state is realized, we can vary $\delta$ to $\delta=0$ and decrease $E$ to $\easync$. To keep $E$ as small as possible during the whole manipulation, we should employ a $\delta$ value close to that of the intersection of the HB and SN curves.
Using the HB curve given by \eqref{eq:origin-hopf-curve} and aSN curve 
given by Eq.~\eqref{eq:SN_lowerbound}, the approximate intersection can be found as
\begin{align}
(\delta, E) \approx \left(\frac{\pi}{2}-2\alpha, 
\frac{{\Omega'}^2 + 4\Lambda^2}{{\Omega'}^2}\right),\label{eq:easynconly-approx}
\end{align}
where
\begin{align}
   \alpha = \arcsin\left[
 \frac{\Omega'}
{\sqrt{{\Omega'}^2+4\Lambda^2}}\right].
   \label{eq:alpha}
   \end{align}
Using this $\delta$ value, 
we can efficiently steer the system into the asynchronous state.

Note that in contrast to the case of the stabilization of the
asynchronous state, the feedback does not vanish when the oscillation-death state
is achieved. 
Moreover, the minimum value of $E$ does
not necessarily imply that the intensity of the feedback $\left|E
f(r)\right|$ is minimized as it also depends on $r$.
Instead, this optimization does minimize 
the possible feedback intensity

\begin{figure*}
    \centering
    \includegraphics[width=17.2cm]{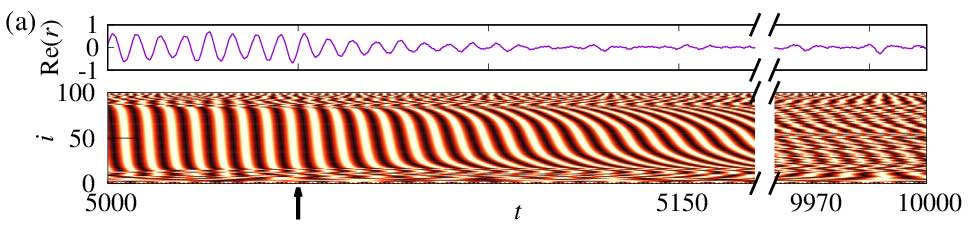}
    \includegraphics[width=17.2cm]{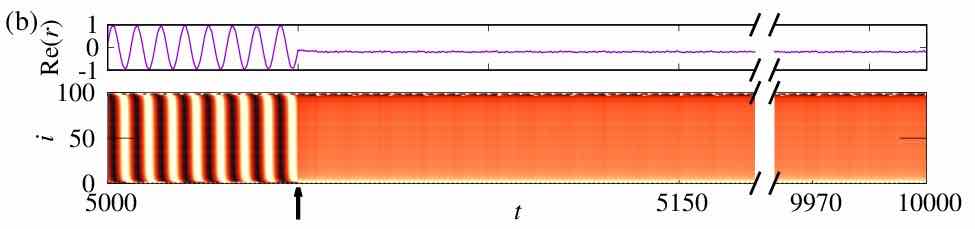}\\
\includegraphics[width=17.2cm]{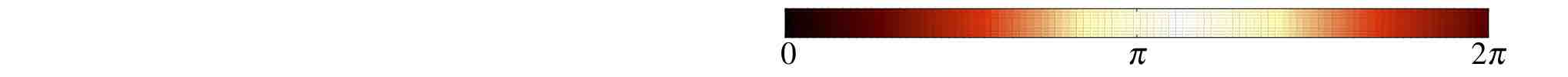}
    \caption{(Color online) Alteration of the dynamics of the $N=100$
 oscillators by the feedback. 
 In each figure, the upper panel displays the time series of $\re{r}$ 
 while the lower panel shows the phases of the oscillators.
 (a) The feedback parameters are set as
 $E=0.3\approx\easync$ and $\delta=0=\dasync$ to stabilize the asynchronous state with small $E$.
 The intrinsic parameters of the oscillators, namely $K$, $\beta$, and $\gamma$, are the same as those in
 Fig. \ref{fig:typical-phase-diagram}(b). (b) The
 feedback parameters are set as $E=1.1\approx \edeath$ and
 $\delta=-\pi/2 \approx \ddeath$ to induce the oscillation-death state with small
 $E$ in the population that has the same parameters as Fig. \ref{fig:typical-phase-diagram}(c).
 }
 \label{fig:sk-dynamics}
\end{figure*}

We perform numerical simulations of Eq.~\eqref{eq:kuramoto-sakaguchi-under-fb} to 
verify whether a near-optimal feedback properly works.
Figure \ref{fig:sk-dynamics} shows the time-series of the collective oscillation 
$\re{r(t)}$ and the individual phases $\theta_i(t)$ before and after the onset of the feedback. 
In Fig. \ref{fig:sk-dynamics}(a), the feedback with the parameters
$E=0.3\approx \easync$ and $\delta =0 = \dasync$ is applied at $t=5050$,
as marked by the black arrow. 
Upon the onset of the feedback, the population begins to be
desynchronized,  
and $r$ decreases with time.
In Fig.\ref{fig:sk-dynamics}(b), the feedback
parameters are chosen such that the oscillation-death state is induced
with small feedback strength: $E=1.10 \approx \edeath$ and $\delta=-\pi/2\approx \ddeath$.
The figure indicates that the oscillations of the individual oscillators and
the mean field terminate immediately because of the feedback.

\section{Investigation on the robustness of the effect of feedback using a different model of oscillators}
\label{sec:different-model}
Our analyses thus far are based on the model given by 
Eq. \eqref{eq:kuramoto-sakaguchi-under-fb}.
As presented in Appendix \ref{sec:derivation}, 
this model is derived from a general class of coupled oscillator models. 
However, we assumed that coupling, inhomogeneity, and feedback
are sufficiently weak to employ averaging approximations and 
that the functions contain only the first harmonics. 
Furthermore, we assumed the natural frequencies to obey 
Lorentzian distribution to obtain 
the reduced dynamical equation given in Eq.~\eqref{eq:OA}.
Here, to exemplify the robustness of the results to the violation of these
assumptions, we provide numerical results for a model with the form given by Eq.~\eqref{eq:winfree}.
Specifically, we consider
\begin{align}
    \dot{\theta_i} = 1 + \mu_i \cos \theta_i 
    + \frac{K}{N}Z_v\left(\theta_i\right) \sum_{j=1}^N V\left(\theta_j\right)
    + E Z_f\left(\theta_i\right) f\left(\bm \theta\right). \label{eq:asmodel}
\end{align}
We adopt the pulse-like signal $V\left(\theta\right)=\nu_n
\left(1+\cos\theta\right)^n$ used in \cite{Ariaratnam2001},
where $n$ is the parameter on the sharpness of $V$, and 
$\nu_n = 2^n(n!)^2/(2n)!$ normalizes $\int_0^{2\pi} V\left(\theta\right)d\theta$
to be $2\pi$.  
We set $n=10$. 
The phase sensitivity functions $Z_v$ and $Z_f$ are
chosen to weakly include the second Fourier mode:
\begin{align}
    Z_v \left(\theta\right) &= -\sin\theta + 0.2 \sin 2\theta,\\
 Z_f\left(\theta\right) &= \sin\theta + 0.2 \cos 2\theta.
\end{align}
Finally, $\mu_i$ is drawn from Gaussian distribution with mean $0$ and
standard deviation $0.1$.

\begin{figure}
    \centering
    \includegraphics[width=8.6cm]{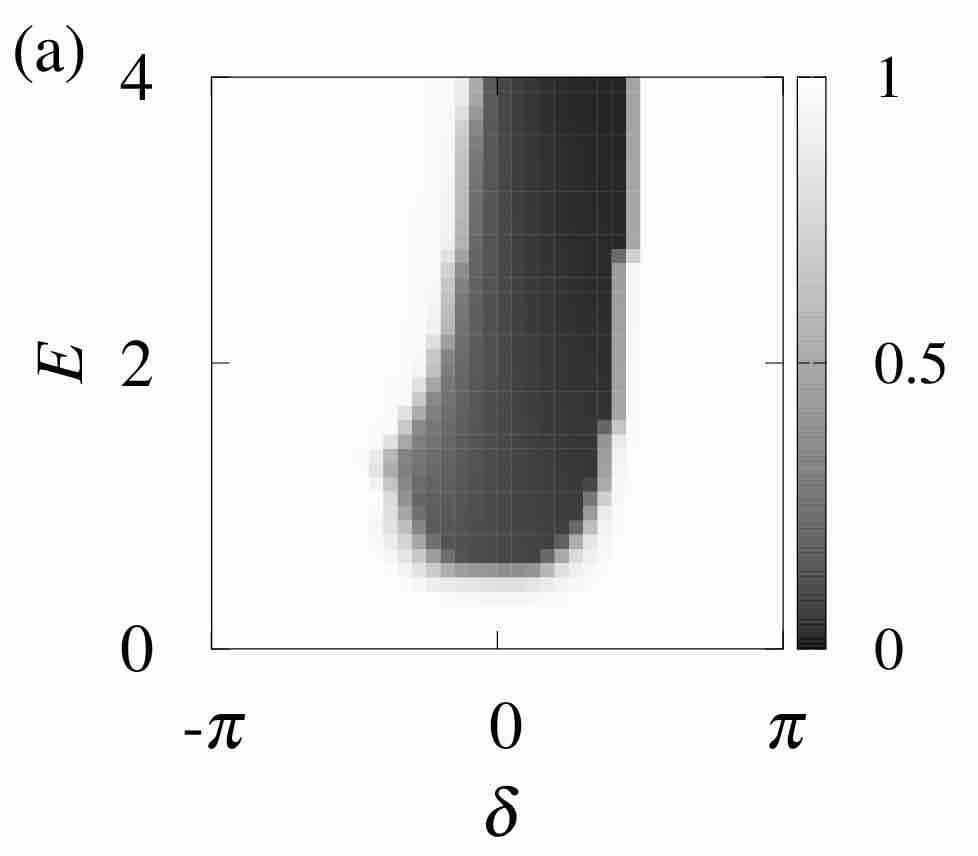}
    \includegraphics[width=8.6cm]{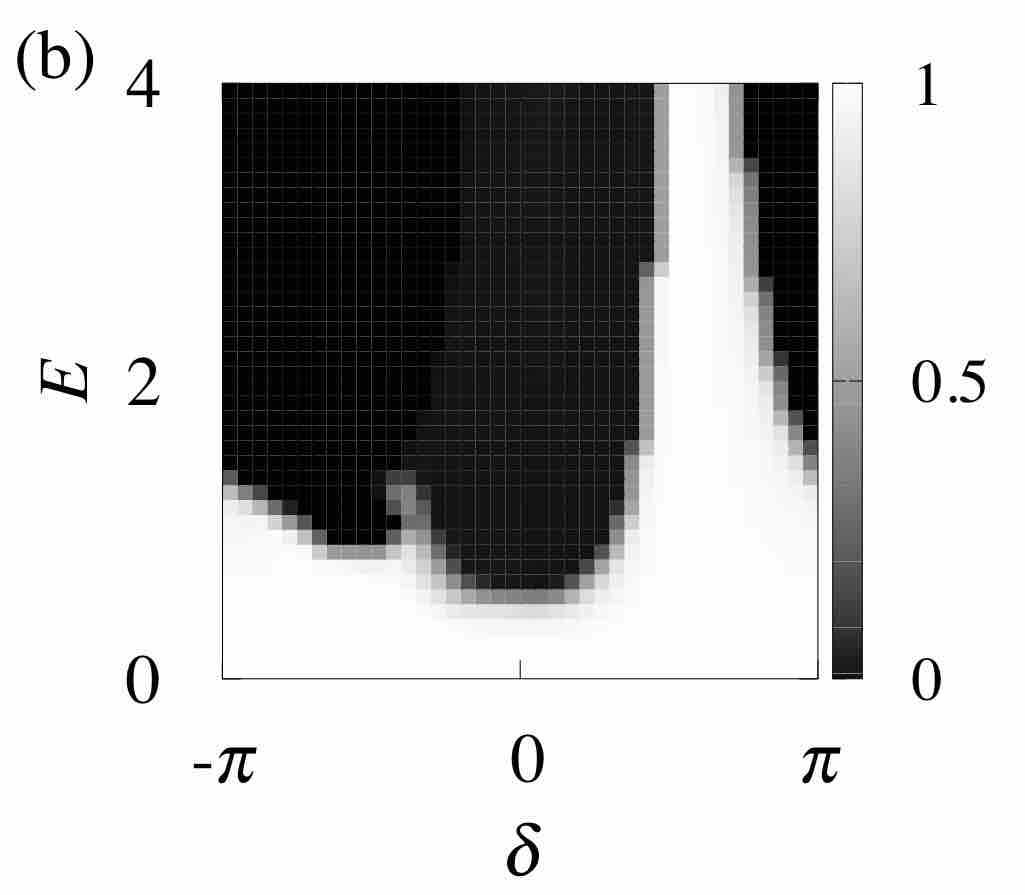}
    \caption{Long-time average of (a) $R$ and (b) $\zeta$ in the system
 of pulse-coupled oscillators under the feedback described by Eq. \eqref{eq:asmodel}. Initial conditions are given by $\theta_i\left(0\right) = 2\pi (i-1)/N$.}
    \label{fig:as-R-zeta}
\end{figure}

Numerical simulation of Eq. \eqref{eq:asmodel} is conducted to calculate 
$\langle R \rangle$ and $\langle \zeta \rangle$, which are shown in
Fig. \ref{fig:as-R-zeta}(a) and (b), respectively.
These figures qualitatively agree with 
Fig.~\ref{fig:2000-ks-RandZ}(a) and 
(d),
suggesting the robustness of the results.

\section{Conclusion and discussion}\label{discussion}
Motivated from the wide range of the applications of synchronization 
control, 
   we analyzed an inhomogeneous population of phase oscillators 
   exposed to global feedback.
Detailed phase diagrams of the collective state are obtained 
based on the bifurcation analysis of the macroscopic equation 
derived using the Ott--Antonsen theory. 
The diagram displayed three types of macroscopic states, namely, 
the synchronously oscillating state, the asynchronous state, and 
the oscillation-death state. 
Exact and approximate optimizations of the feedback parameters 
for steering a synchronously oscillating population into the asynchronous 
or the oscillation-death state with minimum feedback strength are also presented.
Although we assumed several conditions such as the weakness of 
the coupling and the feedback in the derivation of the model equation, 
the numerical investigation in Sec. V 
demonstrates that our results do not change qualitatively
even when some of the assumptions are violated to some extent.

Herein, we focused on linear feedback $F$ given by Eq.~\eqref{eq:feedback}, and
our extensive analysis revealed its utility for synchronization control. 
Linear feedback can be regarded as a basic methodology, 
and our study is expected to serve as a 
benchmark when more sophisticated feedback is to be designed.
A natural extension is to make the feedback function $F$ nonlinear in $R$. 
Although it will not change the linear stability of the asynchronous state, it may 
alter the stability and the existence of other states 
in addition to the amplitude of the collective oscillation\cite{Goldobin2006}
A class of nonlinear feedback has been proposed in \cite{Popovych2005} 
for reducing the amplitude of the collective oscillation, 
and further investigated in successive studies
\cite{Popovych2006,Popovych2018,Goldobin2006,Zhang2019, Popovych2008}. 
In \cite{Kiss2007, Kori2008}, the authors showed that a class of delayed nonlinear
feedback can stabilize complex dynamical states 
including a type of desynchronized state and 
demonstrated its ability to control electrochemical oscillators. 
Furthermore, regarding DBS, smooth feedback may not fulfill safety requirements\cite{Merrill2005}. 
In \cite{Popovych2017}, 
the authors compare two types of feedback: a smooth feedback and 
a series of pulses that are amplified according to the smooth feedback. 
They found that the pulses have a similar desynchronizing effect to the smooth feedback.
The same approach might be applicable to the feedback studied in this article.

Our theoretical results can be demonstrated in some experimental systems.
Recently, techniques for inferring phase dynamics 
and their interactions from observed oscillatory signals have been developed, 
and they have been utilized in experimental studies 
for predicting and controlling the dynamics of the oscillators
\cite{Kiss2005, Kori2014, Stankovski2017, Pikovsky2001book,
Iatsenko2016}. 
These experiments have revealed various coupling and phase sensitivity functions; 
in some systems, the first Fourier component is dominant,
while in other systems, some of the higher and the constant components 
are also prominent. 
In the former case, our analytical results may be verified quantitatively 
by estimating the values of $\omega_i, K,$ and $\beta$ in Eq.~\eqref{eq:kuramoto-sakaguchi-under-fb-sum}
and implementing the global feedback loop. 
The feedback parameters $\delta$ and $E$ can be tuned 
if $R$ and $\Theta$ can be inferred online or two outputs from individual oscillators are available, 
as detailed in Appendix \ref{sec:derivation}.

Finally, we remark on the limitation of the current study. it should be noted that our results are based on phase oscillator models.
A qualitatively different phase diagram may be obtained 
for limit-cycle oscillators whose amplitudes considerably deviate
 from that of the unperturbed periodic orbit.
Therefore, to understand the effect of  
a large amplitude deviation on synchronization control, 
it is important to investigate models of limit-cycle oscillators and compare them with our results.

\begin{acknowledgments}
The authors thank Kei-Ichi Ueda for helpful discussions on numerical
bifurcation analysis. H.K. acknowledges the financial support from JSPS
 KAKENHI Grant No. 18K11464. 
\end{acknowledgments}

H.K. conceived the study and supervised the project. 
A.O. performed the analytical and numerical investigations. 
A.O. wrote the manuscript with support from H.K..

\appendix

\section{Derivation of our model given in Eq.~\eqref{eq:kuramoto-sakaguchi-under-fb-sum-tilde}} \label{sec:derivation}
We consider a general class of oscillator network given by
\begin{align}
 \dot{\bm{x}}_i=\bm{u}_i \left(\bm{x}_i;p_i, q_i\right),\label{eq:original_oscillation}
\end{align}
where $\bm x_i=(x_i, y_i,\ldots)$ and $\bm u_i$ $(i=1,\ldots, N)$ are 
the state and the local vector field of the $i$th oscillator, respectively, 
and $p_i$ and $q_i$ are parameters. 
The interactions and external forcing are assumed to be given through 
variations in the parameters as
$p_i=p_0+\Delta p_i$ and $q_i=q_0+\Delta q_i$,  
where $p_0$ and $q_0$ are the parameter values common 
for all the oscillators and $\Delta p_i$ and $\Delta q_i$  
describe perturbations.
When we consider global coupling and feedback, the perturbations may be given as
\begin{align}
 \Delta p_i &= \frac{K'}{N} \sum_{j=1}^N v(\bm x_i, \bm x_j),\\
 \Delta q_i &= \Delta q \equiv  E' f(\bm x_1, \ldots, \bm x_N),
\end{align}
where $K'$ and $E'$ are the strengths of coupling and feedback, respectively, and $v$ and $f$ are the functions describing coupling and feedback, respectively.

We follow the standard procedure of the phase reduction 
to obtain the corresponding phase description to Eq.~\eqref{eq:original_oscillation}
\cite{Kuramoto1984, Kobayashi2009}.
Let us introduce 
$\Delta \bm u_i(\bm x_i;p_i,q_i) = 
\bm u_i(\bm x_i; p_i, q_i) - \bm u(\bm x_i;p_i,q_i)  $ 
for $i=1,\ldots,N$, which describes inhomogeneity 
in inherent oscillator properties. 
We assume that the unperturbed system, 
i.e., 
\begin{align}
  \dot{\bm x}=\bm u(\bm x; p_0, q_0), \label{eq:unperturbed}
\end{align}
has a stable limit cycle $\bm x^*(t)$. 
The phase of the unperturbed system is defined as a scalar field 
$\Phi(\bm x)$ for the basin of attraction for the limit cycle such 
that the contour of $\Phi(\bm x)$ describe the isochron of the 
unperturbed system\cite{Winfree1967, Kuramoto1984}.
Using this scalar field, the phase of the $i$th oscillator is defined 
as $\theta_i=\Phi(\bm x_i(t))$ ($i=1,\ldots,N$). 

We assume that the orbital stability of the cycle $\bm x^*(t)$ 
 in the unperturbed system given by Eq. \eqref{eq:unperturbed}
is sufficiently higher than perturbation strength.
Then, to the lowest order in perturbations strengths, 
each phase obeys
\begin{align}
 \dot \theta_i = \omega + \bm Z(\theta_i) \cdot \bm U_i(\theta_i) + \frac{K'}{N} \sum_{j=1}^N
 Z_v(\theta_i) V(\theta_i, \theta_j) + E' Z_f(\theta_i) F(\bm \theta), 
 \label{eq:winfree}
\end{align}
where $\omega$ is the natural frequency of the limit cycle 
$\bm x^*$; $\bm U_i, V, F$ are the parametric representations of 
$\Delta \bm u_i, v, f$ on the limit-cycle $\bm x^*$ in terms of the phases, 
respectively; 
and $\bm Z, Z_v, Z_f$ are the phase sensitivity functions, 
which can be expressed in terms of the derivatives of 
$\bm u(\bm x; p, q)$ and $\Phi\left( \bm{x}\right)$. 
All the functions are $2\pi$-periodic in each argument.

When the magnitudes of the perturbation terms, i.e., the second to fourth terms of the right-hand side in Eq.~\eqref{eq:winfree}, are sufficiently small compared to $\omega$, we may further simplify the equation using an averaging approximation. The resultant equation is
\begin{align}
 \dot \theta_i = \omega + \Delta \omega_i + \frac{K'}{N} \sum_{j=1}^N
 \Gamma_v(\theta_i- \theta_j) + \frac{E'}{N} \sum_{j=1}^N \Gamma_f(\theta_i- \theta_j), \label{eq:kuramoto}
\end{align}
where the constant $\Delta \omega_i$ and the functions $\Gamma_v$ and $\Gamma_f$ 
can be expressed in terms of the functions appearing in Eq.~\eqref{eq:winfree}.
If the last term in Eq.~\eqref{eq:winfree} is not 
very small compared to $\omega$, we can still average the other terms to obtain 
\begin{align}
 \dot \theta_i = \omega + \Delta \omega_i + \frac{K'}{N} \sum_{j=1}^N
 \Gamma_v(\theta_i- \theta_j) + E' Z_f(\theta_i) F(\bm \theta). \label{eq:kuramoto2}
\end{align}
In Eq.~\eqref{eq:kuramoto2}, we may consider larger $E'$ values than in
Eq.~\eqref{eq:kuramoto}, 
which is the reason why we employ this type of model in this work.

Our model given in Eq.~\eqref{eq:kuramoto-sakaguchi-under-fb-sum-tilde} 
is a version of Eq.~\eqref{eq:kuramoto2}, 
where we assume that only the first harmonics are present 
in all the functions appearing in Eq.~\eqref{eq:kuramoto2}; 
i.e., $\Gamma_v$, $Z_f$ and $F$. 
This assumption is valid when we consider 
limit-cycle oscillators 
close to a Hopf bifurcation point, 
in which the phase sensitivity and the wave forms 
are well approximated by the functions 
with only the first harmonics and with the constant 
and first harmonics terms, respectively. 
Let us further assume that from each oscillator 
we can observe two quantities, such as $x_j(t)$ and $y_j(t)$. 
    If we denote the trajectory of the unperturbed limit cycle $\bm x^*(t)$ 
by $\bm \chi(\Phi)$, i.e., $\bm \chi(\Phi(\bm x^*(t)))=\bm x^*(t)$, 
with its elements being $\bm \chi=(\chi_x,\chi_y,\ldots)$, 
the variation of $\chi_x\left(\theta\right)$ and $\chi_y\left(\theta\right)$ 
are almost sinusoidal near the Hopf bifurcation point. Therefore, 
the unperturbed waveforms can be denoted by 
$\chi_x(\theta) \simeq \bar{\chi}_x +  A_x \cos(\theta-\delta_x)$ and 
$\chi_y(\theta) \simeq \bar{\chi}_y + A_y \cos(\theta-\delta_y)$, 
where $\bar{\chi}_x$ and $\bar{\chi}_y$ are the average of 
$\chi_x(\theta)$ and $\chi_y(\theta)$, respectively, 
$A_x$ and $A_y$ are the oscillation amplitudes, 
and $\delta_x$ and $\delta_y$ are the phase offsets in the waveforms.
We then give the feedback function $f$ as 
\begin{align}
 f=\sum_{j=1}^{N} \left[a \frac{x_j - \bar x_j}{A_x}+ b \frac{y_j -\bar y_j}{A_y}\right],
\label{eq:feedback_raw}
\end{align}
where $\bar x_j$ and $\bar y_j$ denote the average values of $x_j$ and $y_j$, respectively; 
and $a$ and $b$ are our control parameters. 
In the lowest order phase description, Eq.~\eqref{eq:feedback_raw} results in
\begin{align}
 F(\bm{\theta}) = \sum_{j=1}^N \left[a \cos(\theta_j - \delta_x) + b \cos(\theta_j-\delta_y)
 \right].
\label{eq:Fab}
\end{align}
We can further transform \eqref{eq:Fab} to
\begin{align}
 F(\bm{\theta}) &=\mathcal{E} \sum_{j=1}^N\cos\left( \theta_j - \delta\right)\\
 &= \mathcal{E}R \cos\left( \Theta - \delta\right),
\end{align}
where 
\begin{align}
 \mathcal{E}&= \left[\left(a\cos\delta_x + b
   \cos\delta_y\right)^2+\left(a\sin\delta_x + b \sin\delta_y\right)^2
   \right]^{1/2}, \\
  \tan\delta&=\frac{a\sin\delta_x + b \sin\delta_y}{a\cos\delta_x + b
   \cos\delta_y}.
  \end{align}
Therefore, we can give arbitrary $\mathcal{E}$
and $\delta$ values by appropriately assigning $a$ and $b$ values.
Because $\tilde{E}$ in
Eq.~\eqref{eq:kuramoto-sakaguchi-under-fb-sum-tilde} is given by $\tilde{E}=E' \mathcal{E}$,
we can also give an arbitrary $\tilde{E}$ value.

\section{Classification of the zero-eigenvalue bifurcation at the origin}
\label{sec:zeroeigen-origin}
The codimention-one bifurcation involving zero-eigenvalue at the origin
is limited 
to the pitchfork bifurcation. One possible approach for this 
is to consider the facts that the saddle--node bifurcation may 
not occur because the constant solution $r=0$ may not vanish in Eq.~\eqref{eq:OA} 
and the pitchfork bifurcation rather than the transcritical bifurcation occurs 
because the symmetry of Eq.~\eqref{eq:OA} implies that the emergence of a constant 
solution $r=r^* \neq 0$ must be accompanied with the emergence of $r=-r^*$ as well.

An alternative way is to perform the center manifold reduction\cite{guckenheimer}, 
which will clarify that the bifurcation is actually the pitchfork one 
and whether the bifurcation is super- or sub-critical.
Let the bifurcation parameter be 
$\mu = E - \epf$, where $\epf$ is given by 
 Eq.~\eqref{eq:origin-pf-curve}.
Inserting $r = u + \ii v$ and $E = \epf + \mu$ into Eq.~\eqref{eq:OA}, 
we obtain
\begin{align}
   \begin{pmatrix}
      \dot{u} \\ \dot{v}
   \end{pmatrix}
   =
   \begin{pmatrix}
      \Lambda-\frac{\epf \cos\delta}{2} & -\Omega-\frac{\epf \sin\delta}{2}\\
      \Omega & \Lambda
   \end{pmatrix}
   \begin{pmatrix}
      u \\ v
   \end{pmatrix}
   +
   \begin{pmatrix}
      p\left(u,v\right) \\
      q\left(u,v\right)
   \end{pmatrix},
\end{align}
where
\begin{align}
 p(u,v)&=-\frac{\mu}{2} (u \cos \delta +v \sin \delta ) 
 +\frac{\epf+\mu }{2}  \left(u^2-v^2\right) (u \cos \delta +v \sin
 \delta ) \notag \\
 &~~~~-\frac{K}{2}\left(u^2+v^2\right) \left( u \cos \beta + v \sin \beta \right),\\
   q\left(u,v\right) &= u^2 v \left[(\epf+\mu ) \cos \delta-\gamma
 -\Lambda \right]+u v^2 \left[(\epf+\mu ) \sin \delta+\Omega -1\right] \notag\\
   &~~~~+u^3 (\Omega -1)+v^3 (-\gamma -\Lambda ).
\end{align}
To reduce the system, let us transform the variables as follows:
\begin{align}
   \begin{pmatrix}
      u \\ v
   \end{pmatrix}
   =
   \begin{pmatrix}  
      -\frac{\Lambda }{\Omega } & \frac{\Lambda  \sin \delta+\Omega  \cos \delta}{\Omega  \sin \delta-\Lambda  \cos \delta} \\
 1 & 1 \\
   \end{pmatrix}
   \begin{pmatrix}
      \hat{u} \\ \hat{v}
   \end{pmatrix},
\end{align}
which yields
\begin{align}
   \begin{pmatrix}
      \dot{\hat{u}} \\ \dot{\hat{v}}
   \end{pmatrix}
   =
   \begin{pmatrix}
      0 & 0 \\
      0 & \hat{\lambda}
   \end{pmatrix}
   \begin{pmatrix}
      \hat{u} \\ \hat{v}
   \end{pmatrix}+
   \begin{pmatrix}
      a_1 \mu \hat{u} + a_2 \mu \hat{v} + a_3 \hat{u}^3 + O\left(v^3, u v^2, u^2 v\right)\\
      b \mu \hat{u} + O\left(\mu \hat{v}, \hat{v}^3, \hat{u} \hat{v}^2, \hat{u}^2 \hat{v}, \hat{u}^3\right)
   \end{pmatrix}
\end{align}
where 
\begin{align}
   \hat{\lambda} &= \frac{\cos \delta \left(\Lambda ^2-\Omega ^2\right)-2 \Lambda  \Omega  \sin \delta}{\Lambda  \cos \delta-\Omega  \sin \delta},\\
   a_1 &= \frac{(\Omega  \sin \delta-\Lambda  \cos \delta)^2}
   {2 \left[\left(\Omega^2 - \Lambda ^2\right) \cos \delta+2 \Lambda  \Omega  \sin \delta\right]},\\
   a_2 &= \frac{\Omega ^2 }
 {2 \left[\left(\Omega^2 - \Lambda ^2\right) \cos \delta+2 \Lambda  \Omega  \sin \delta\right]},\\
   a_3 &= \frac{\left(\Lambda ^2+\Omega ^2\right) 
   \left\{ \left(-2 \gamma  \Lambda  \Omega +\Lambda ^2-\Omega ^2\right)\sin \delta
   + \left[\gamma  \left(\Lambda ^2-\Omega ^2\right)+2 \Lambda  \Omega \right]\cos \delta\right\}}
   {\Omega ^2 \left[ \left(\Omega
   ^2-\Lambda ^2\right)\cos \delta+2 \Lambda  \Omega  \sin \delta\right]},\\
 b_1&=\frac{(\Lambda  \cos \delta -\Omega  \sin \delta )^2}{2  \left(\Lambda ^2-\Omega ^2\right)\cos \delta -4 \Lambda  \Omega  \sin (\delta )}.
\end{align}
The center manifold $\hat{v} = c\left(\hat{u}, \mu\right)$ up to the second order is given by
\begin{align}
   c\left(\hat{u}, \mu\right) = -\frac{b}{\hat{\lambda}} \mu\hat{u} + 
 O\left( \hat{u}\mu^2, \hat{u}^3\right).
\end{align}
On this center manifold, the dynamics is reduced to
\begin{align}
   \dot{\hat{u}} = a_1 \mu \left(1 - \frac{a_2 b}{\hat{\lambda}a_1}\mu\right) \hat{u}
   + a_3 \hat{u}^3 + O\left(\mu \hat{u}^3, \mu^3 u\right).\label{eq:cm-pf}
\end{align}
Equation~\eqref{eq:cm-pf} implies that the pitchfork bifurcation occurs at $\mu=0$.
The origin changes its stability through this bifurcation when $\hat{\lambda}<0$.
It is supercritical for $a_3 < 0$ and subcritical for $a_3 > 0$.

Below, we show that the bifurcation is subcritical when $
\hat{\lambda}<0$, $\Omega>0$, 
and $\epf$ is sufficiently small compared to $\Omega/\gamma$.
Because $E \geq 0$, and the numerator of $\epf$ is positive, 
the denominator of $\epf$ is also positive:
\begin{align}
   \Lambda \cos \delta - \Omega \sin \delta > 0, \label{eq:ec>0}
\end{align}
which implies that the denominator of 
$\hat{\lambda}$ is positive.
Thus, if $\hat{\lambda} < 0$, 
the numerator of $\hat{\lambda}$ is negative, i.e., 
\begin{align}
   \left(\Lambda^2 - \Omega^2\right)\cos \delta - 2 \Lambda \Omega \sin \delta < 0.
   \label{eq:lamhat<0}
\end{align}
Under this condition, $a_3$ is positive if
\begin{align}
   &
   \left(\gamma  \left(\Lambda ^2-\Omega ^2\right)+2 \Lambda  \Omega \right)
   \cos \delta 
 >  
    \left(2 \gamma  \Lambda  \Omega -\Lambda ^2+\Omega ^2\right)
    \sin \delta\\
   \iff
   &\Omega \left(\Lambda \cos \delta - \Omega \sin \delta\right)
 >  
 -\Lambda\left(\Lambda \sin \delta + \Omega \cos \delta\right) +
   \gamma \left[2 \Lambda \Omega \sin \delta - \left(\Lambda^2 - \Omega^2\right)\cos \delta\right]
   \label{eq:pf-spr-cond}
\end{align}

We first consider the case  
of $\Lambda \geq 0$. 
Then, the first term of the right-hand side of
Eq.~\eqref{eq:pf-spr-cond} is negative because 
Eqs.~\eqref{eq:ec>0} and \eqref{eq:lamhat<0} yield
\begin{align}
   0 \leq \Lambda\left(\Lambda\cos \delta - \Omega \sin \delta\right) < \Omega \left(\Lambda\sin \delta + \Omega\cos \delta\right).
   \label{eq:pf-1st-term-negative}
\end{align}
As for the second term, noting that $\left|\cos\delta\right|,~\left| \sin \delta \right| \leq 1$,  
we obtain
\begin{align}
   2 \Lambda \Omega \sin \delta - \left(\Lambda^2 - \Omega^2\right) \cos \delta
   \leq
   2 \Lambda \Omega + \Lambda^2 + \Omega^2 
   \leq
   2\left(\Lambda^2 + \Omega^2\right)
   \label{eq:pf-2nd-term-limit}
\end{align}
Equations~\eqref{eq:pf-spr-cond},~\eqref{eq:pf-1st-term-negative},
and~\eqref{eq:pf-2nd-term-limit}
yield the following sufficient condition for $a_3$ to be positive:
\begin{align}
   \Omega \left( \Lambda \cos \delta - \Omega \sin \delta\right)
 >  
 2\gamma \left(\Lambda^2 + \Omega^2 \right),
\end{align}
which is equivalent to  
\begin{align}
   \epf = \frac{2\left(\Lambda^2 + \Omega^2\right)}{\Lambda\cos \delta -
 \Omega\sin \delta} 
<  
 \frac{\Omega}{\gamma}. 
   \label{eq:pf-ec-cond1}
\end{align}
Therefore, for $\Lambda \geq 0$, the bifurcation is subcritical when
$\epf$ is sufficiently small compared to $\Omega/\gamma$.

We next consider the case 
of $\Lambda < 0$.
If $\Lambda\sin\delta + \Omega\cos \delta <0$ holds, 
we can derive Eq.~\eqref{eq:pf-ec-cond1} in the same manner as the case of $\Lambda\geq 0$.
When $\Lambda\sin\delta + \Omega\cos \delta \geq 0$, we evaluate the right-hand side 
of Eq.~\eqref{eq:pf-spr-cond} as follows.
As $\left|\Lambda\right| < \gamma$ holds for $\Lambda < 0$, 
\begin{align}
   -\Lambda\left(\Lambda\sin\delta + \Omega\cos \delta\right)
   <
   \gamma\left(\left|\Lambda\right| + \Omega\right).
\end{align}
This equation and Eqs.~\eqref{eq:pf-spr-cond} and \eqref{eq:pf-2nd-term-limit} yield 
the sufficient condition for the subcritical bifurcation
\begin{align}
   \Omega\left(\Lambda\cos\delta-\Omega\sin \delta\right)
   >  
   2\gamma\left(\Lambda^2 + \Omega^2\right)\left(1 + \frac{\left|\Lambda\right| + \Omega}{2\left(\Lambda^2 + \Omega^2\right)}\right),
\end{align}
which is equivalent to 
\begin{align}
   \epf  <
 \frac{\Omega}{\gamma\left(1 + \frac{\left|\Lambda\right| + \Omega}{2\left(\Lambda^2 + \Omega^2\right)}\right)}.
   \label{eq:pf-ec-cond2}
\end{align}
From Eq.\eqref{eq:pf-ec-cond1} and Eq.\eqref{eq:pf-ec-cond2}, 
it is shown that the pitchfork bifurcation is subcritical if $\epf$ is sufficiently 
small compared to $\Omega/\gamma$.

Near the parameter regions considered in
Fig.~\ref{fig:typical-phase-diagram}, $\epf$ is small enough, 
and hence the pitchfork bifurcation involving a stable fixed point is subcritical.

\section{Weakly nonlinear analysis of the Hopf bifurcation}
\label{sec:weakly-nlin-analysis}
We show below that the Hopf bifurcation at the origin is always supercritical.
Substituting $r = u + \ii v$ and the value of the feedback strength
at the bifurcation point, given by Eq.~\eqref{eq:origin-hopf-curve}, into Eq.~\eqref{eq:OA},
we have
\begin{align}  
 \left(
\begin{array}{c}
 \dot{u}\\
 \dot{v}
\end{array}
\right)=
\left(
\begin{array}{cc}
 -\Lambda  & -2 \Lambda  \tan \delta-\Omega  \\
 \Omega  & \Lambda  \\
\end{array}
\right)
 \left(
\begin{array}{c}
 u\\
 v
\end{array}
\right)+
\left(
\begin{array}{c}
g\left(u,v\right)\\
 h\left(u,v\right)
\end{array}
\right),
\end{align}
where 
\begin{align}
g(x,y)&=2 \Lambda  \left(x^2-y^2\right) (x+y \tan \delta)-\frac{K}{2} \left(x^2+y^2\right) (x \cos \beta
   +y \sin \beta),\\
h(x,y)&=\frac{K}{2} \left(x^2+y^2\right) (x \sin \beta-y \cos \beta)+4 \Lambda  x y (x+y \tan \delta   )
\end{align}
To simplify the calculation, we change the coordinates to  
\begin{align}
 \left(
\begin{array}{c}
 \tilde{u}\\
\tilde{v}
\end{array}
\right)=
\left(
\begin{array}{cc}
 \Omega/\tilde{\omega}&\Lambda/\tilde{\omega}\\
0 & 1
\end{array}
\right)
\left(
\begin{array}{c}
 u\\
v
\end{array}
\right),
\end{align}
where $\tilde{\omega}=\sqrt{-\Lambda^2 + \Omega^2 + 2 \Lambda\Omega \tan
\delta}$.
Then, $\tilde{u}$ and $\tilde{v}$ obey
\begin{align}
\label{eq:tilde-u-v}
 \left(
\begin{array}{c}
 \dot{\tilde{u}}\\
 \dot{\tilde{v}}
\end{array}
\right)=
\left(
\begin{array}{cc}
 0& -\tilde{\omega} \\
\tilde{\omega} & 0
\end{array}
\right)
\left(
\begin{array}{c}
 \tilde{u}\\
\tilde{v}
\end{array}
\right)+
\left(
\begin{array}{c}
 \tilde{g}\left(\tilde{u},\tilde{v}\right)\\
 \tilde{h}\left(\tilde{u},\tilde{v}\right)
\end{array}
\right),
\end{align}
where 
\begin{align}
\left(
\begin{array}{c}
 \tilde{g}\left(\tilde{u},\tilde{v}\right)\\
 \tilde{h}\left(\tilde{u},\tilde{v}\right)
\end{array}
\right)=
\left(
\begin{array}{cc}
 \Omega/\tilde{\omega}&\Lambda/\tilde{\omega}\\
0 & 1
\end{array}
\right)\left(
\begin{array}{c}
 g\left(u\left(\tilde{u},\tilde{v}
\right),v\left(\tilde{u},\tilde{v}
\right)\right)\\
 h\left(u\left(\tilde{u},\tilde{v}
\right),v\left(\tilde{u},\tilde{v}
\right)\right)
\end{array}
\right).
\end{align}
Then, $z=u + \ii v$ obeys 
\begin{align}
 \dot{z} = \ii \tilde{\omega}z + \tilde{g} + \ii \tilde{h}.
 \label{eq:dzdt}
\end{align}
Using a near-identity transformation, Equation \eqref{eq:dzdt} is
further reduced 
to  
the following normal form of the Hopf bifurcation \cite{guckenheimer}
\begin{align}
 \dot{w} = \ii \tilde{\omega} w + d \left|w\right|^2 \bar{w} +
 O\left(\left|w\right|^5\right),
\end{align}
where $w, d \in \C$, and 
\begin{align}
 \re{d} = -\gamma\left(1 + \frac{\Lambda}{\Omega}\tan\delta\right).\label{eq:re(c)}
\end{align}

The bifurcation is supercritical when $\re{d} < 0$. 
Noting that $\gamma > 0$, we may further show  
$\re{d} < 0$ as follows.  
When $\frac{\Lambda}{\Omega}\tan\delta \geq 0$, it is obvious that
$\re{d}< 0$ from Eq. \eqref{eq:re(c)}. When
$\frac{\Lambda}{\Omega}\tan\delta < 0$, the following inequality holds:
\begin{align}
 1 + \frac{\Lambda}{\Omega}\tan\delta > 
1 + 2\frac{\Lambda}{\Omega}\tan \delta.
\label{eq:tand>2tand}
\end{align}
Moreover, Eq.~\eqref{eq:origin-hopf-curve-domain} implies that the Hopf
bifurcation at the origin occurs only when
\begin{align}
 -\Lambda^2 + \Omega^2 + 2 \Lambda \Omega \tan \delta > 0.
\label{eq:wnahopf-hopfcond}
\end{align}
From Eqs.~\eqref{eq:tand>2tand} and \eqref{eq:wnahopf-hopfcond}, we have
\begin{align}
 1 + \frac{\Lambda}{\Omega}\tan\delta >  \left(\frac{\Lambda}{\Omega}\right)^2 > 0
\end{align}
and hence $\re{d} < 0$. Therefore, the Hopf bifurcation at the origin is
supercritical for any parameter values.

\section{Derivation of the sufficient condition for a nonzero fixed point}
\label{sec:death-sufficient}
   Here, under the conditions $K>2\kc$ and $\Omega' = 1 + K\sin\beta>0$, 
we derive  
Eqs.~\eqref{eq:death-sufficient_lower} and \eqref{eq:death-sufficient_higher}, 
i.e., a sufficient condition for Eq.~\eqref{eq:sk:oa:polar} 
to have a nonzero fixed point. 
Note that 
   an intersection of the two nullclines
\begin{align}
   -\gamma + \left(1-R^2\right)
   \left[\frac{K}{2}\cos\beta- \frac{E}{4}\left(\cos\left(\delta-2\Theta\right)+\cos\delta\right)\right]=0
   \label{eq:nullcline-R}
\end{align}
and
\begin{align}
   1+\left(1+R ^2\right) \left[\frac{K}{2}  \sin
   \beta+\frac{E}{4}\left(\sin \delta -\sin \left(\delta -2 \Theta \right)\right)\right]=0  
   \label{eq:nullcline-Th}
\end{align}
gives a nonzero fixed point.
Thus there exists a fixed point $\left( R^*, \Theta^*\right)$ such that 
$\rl < R^* < 1$ if both of the following conditions are satisfied: 
(i) the nullcline given by Eq.~\eqref{eq:nullcline-R} 
is defined for any $\Theta$, and the value of $R$ on the nullcline always satisfies 
 $\rl < R^* < 1$, and (ii) the nullcline given by Eq.~\eqref{eq:nullcline-Th} 
 passes through the region $\rl < R^* < 1$ on the $R$-$\Theta$ plane.

The first condition is 
equivalent to that the following inequality holds for any $\Theta$:
\begin{align}
   {\rl}^2 < 
   1 -    \frac{4\gamma}
   {2 K \cos \beta - 
   E \left[\cos\left(\delta - 2 \Theta\right)+ \cos \delta\right]} < 1,
\end{align}
which is satisfied when 
\begin{align}
   E < 
   \frac{2 K \cos \beta - 4 \gamma\left(1-{\rl}^2\right)^{-1}}
   {1 + \cos \delta}.
   \label{eq:sufficient-upperE}
\end{align}

Next we discuss the second condition. 
Equation \eqref{eq:nullcline-Th} yields 
$R^2 = -1 + S\left(\sin \left(\delta - 2 \Theta\right)\right)^{-1}$, where
\begin{align}
   S\left( x \right) 
   = -\frac{K}{2}\sin \beta + \frac{E}{4}\left(x - \sin \delta\right) 
\end{align}
is a monotonically increasing function of $x$. 
Because $\sin\left(\delta - 2 \Theta\right)$ is in the range of
$\left[1,1\right]$, the nullcline given by Eq.~\eqref{eq:nullcline-Th} 
crosses over the region $\rl < R^* < 1$ if 
\begin{align}
   S\left(-1\right)<\frac{1}{2}  \label{eq:s(-1)}
\end{align}
and
\begin{align}
  S\left(1\right)> \left(1 + \rl^2\right)^{-1}. \label{eq:s(1)}
\end{align}
When $\Omega'>0$, the inequality given by \eqref{eq:s(-1)} holds 
for any $E \geq 0$.  
In contrast, inequality \eqref{eq:s(1)} holds when 
\begin{align}
   E > E_{\rm lower} 
   \left[1 + 
   \frac{1-{\rl}^2}
   {\left(1+{\rl}^2\right)\left(1 + K\sin \beta \right)}\right],
   \label{eq:sufficient-lowerE}
\end{align}
where $E_{\rm lower}$ is given by Eq.~\eqref{eq:SN_lowerbound}.

Therefore, if Eqs.~\eqref{eq:sufficient-upperE} 
and \eqref{eq:sufficient-lowerE} are satisfied, 
a nonzero fixed point $\left(R^*, \Theta^*\right)$ that satisfies 
 $\rl < R^* < 1$ exists. As a special case, 
 we obtain 
 Eqs. \eqref{eq:death-sufficient_lower} and
 \eqref{eq:death-sufficient_higher}, respectively, 
 by setting $\rl = \sqrt{1 - \frac{2\kc}{K}}$ in 
Eqs.~\eqref{eq:sufficient-lowerE} and \eqref{eq:sufficient-upperE}. 

We can find a value of $E$ that satisfies 
both of Eqs. \eqref{eq:death-sufficient_lower} and 
\eqref{eq:death-sufficient_higher} when 
\begin{align}
   E_{\rm lower} 
   \left(1 + \eta \right)
   < \frac{K \cos \beta }{1 + \cos \delta},
\end{align}
which is equivalent to
\begin{align}
\frac{2}{1 - \sin \delta }\left[\frac{1}{K - \kc}+ \sin \beta\right]
- \frac{\cos \beta }{1 + \cos \delta }
< 0.
\label{eq:death-sufficient-param}
\end{align}
Equation \eqref{eq:death-sufficient-param} holds when $\beta \simeq 0$ and 
$K$ is sufficiently large for a given value of $\delta$.

\section{Analysis on the codimension-two bifurcation point}
\label{sec:cod2}
By imposing $\tr{L}=\det{L}=0$, we obtain the values of 
the feedback parameters at which the Hopf
bifurcation curve \eqref{eq:origin-hopf-curve} and the pitchfork
bifurcation curve \eqref{eq:origin-pf-curve} meet  as
\begin{align}
 \sin\delta &= 
 \frac{\Omega}{\left|\Omega\right|}\frac{\Lambda^2 - \Omega^2}{\Lambda^2 + \Omega^2}, \label{eq:sindc}\\
 \cos\delta &= 
 \frac{2\Lambda\left|\Omega\right|}{\Lambda^2 + \Omega^2}, 
\label{eq:cosdc}\\
E&=
 \frac{2\left(\Lambda^2 + \Omega^2\right)}{\left|\Omega\right|}. \label{eq:ec}
\end{align}

Substituting $r=u+\ii v$ together with
Eq. \eqref{eq:sindc}--\eqref{eq:ec} into Eq. \eqref{eq:OA}, we have
\begin{align}
 \begin{pmatrix}
  \dot{u} \\ \dot{v}
 \end{pmatrix}
=
\begin{pmatrix}
 -\Lambda  & -\frac{\Lambda ^2}{\Omega } \\
 \Omega  & \Lambda
\end{pmatrix}
\begin{pmatrix}
 u \\ v\\
\end{pmatrix}
+
\begin{pmatrix}
 g_2\left(u,v\right)\\
 h_2\left(u,v\right)
\end{pmatrix},
\end{align}
where
\begin{align}
 g_2\left(u,v\right) &= (\Lambda -\gamma )u^3+\left(1-2
 \Omega +\frac{\Lambda ^2 }{\Omega }\right)u^2v
 - (\gamma +3 \Lambda ) u v^2
+ \left(1 - \frac{\Lambda^2}{\Omega}\right)v^3,\\
 h_2\left(u,v\right) &= (\Omega -1)u^3
- (\gamma -3 \Lambda )u^2v 
+\left( -1 - \Omega + \frac{2\Lambda^2}{\Omega}\right)u v^2
-(\gamma+\Lambda )v^3.
\end{align}
Next, we transform the linear part into the Jordan normal form by the
following change of variables:
\begin{align}
 \begin{pmatrix}
  \check{u} \\ \check{v}
 \end{pmatrix}
=
\begin{pmatrix}
 0 & 1 \\
 \Omega  & \Lambda 
\end{pmatrix}
\begin{pmatrix}
 u \\ v
\end{pmatrix},
\end{align}
which yields
\begin{align}
\label{eq:dot(hat(u))}
 \begin{pmatrix}
  \dot{\check{u}}\\ \dot{\check{v}}
 \end{pmatrix}
=
\begin{pmatrix}
 \check{v} \\ 0 
\end{pmatrix}
+
\begin{pmatrix}
 c_1 \check{u}^3 + c_2 \check{u}^2 \check{v} + c_3 \check{u} \check{v}^2 + c_4 \check{v}^3, \\
 d_1 \check{u}^3 + d_2 \check{u}^2 \check{v} + d_3 \check{u} \check{v}^2 + d_4 \check{v}^3
\end{pmatrix},
\end{align}
where 
\begin{align}
 c_1 &= \left(\Lambda ^2+\Omega ^2\right) (\Lambda
 -\gamma  \Omega )/\Omega ^3,\\
 c_2 &= \left(2 \gamma  \Lambda
\Omega -\Lambda ^2 \Omega -3 \Lambda ^2  -  \Omega
^2-\Omega ^3\right)/\Omega^3,\\
c_3 &= \left(3 \Lambda   -\gamma
\Omega\right)/\Omega^3, \\
c_4 &= \left(-1 +
\Omega\right)/\Omega^3, \\
d_1 &=  \left(\Lambda ^2+\Omega
^2\right)^2/\Omega^3,\\
d_2 &= \left(\Lambda ^2+\Omega ^2\right)
(-\gamma  \Omega -3 \Lambda  )/ \Omega^3,\\
d_3 &= \left[  \left(3 \Lambda ^2+\Omega ^2\right)-2 \Omega  
\left(-\gamma  \Lambda +\Lambda ^2+\Omega ^2\right)\right] /\Omega^3,\\
d_4 &=  \left[-\Lambda -\Omega  (\gamma -2 \Lambda )\right]/\Omega^3.
\end{align}

Equation \eqref{eq:dot(hat(u))} is then reduced to
\begin{align}
 \begin{pmatrix}
  \dot{U} \\ \dot{V}
 \end{pmatrix}
=
\begin{pmatrix}
 V \\ 0
\end{pmatrix}
+
\begin{pmatrix}
  0 \\
d_1 U^3 + \left(3 c_1 + d_2 \right)U^2 V
\end{pmatrix}
+O\left(U^5, U^4V, U^3V^2, U^2 V^3, U V^4, V^5\right)
\end{align}
by the following near-identity transformation
\begin{align}
 \begin{pmatrix}
  \check{u} \\ \check{v}
 \end{pmatrix}
= 
\begin{pmatrix}
 U \\ V
\end{pmatrix}
+
\begin{pmatrix}
 \frac{1}{6}\left(2 c_2+d_3\right)U^3+\frac{1}{2}\left(c_3+d_4
\right) U^2 V + c_4 U V^2\\
-c_1 U^3+\frac{1}{2} d_3 U^2 V+d_4 U V^2
\end{pmatrix}.
\end{align}

The signs of $d_1$ and $d_2':=3c_1 +
d_2=-4\gamma\left(\Lambda^2 + \Omega^2\right)/\Omega^2$ determine the
types of the codimension-one bifurcation that occurs near the
codimension-two bifurcation point\cite{guckenheimer}. 
The heteroclinic bifurcation as well as the Hopf and the pitchfork
bifurcations occurs for $d_1 d_2'< 0$, while the bifurcation involving a
pair of homoclinic orbit occurs for $d_1 d_2' > 0$\cite{guckenheimer}. The former is the 
case for $\Omega>0$, and the latter is the case for $\Omega<0$.

Note that the analysis above only gives the information around 
a specific codimension-two bifurcation point and does not necessarily imply  
that the Hopf bifurcation involving nonzero fixed point  
may not occur for $\Omega > 0$. 
In practice, we numerically observed it for some parameter sets
even when $\Omega > 0$.

\section{Numerical search for limit-cycle solutions of Eq. \eqref{eq:OA}}
\label{sec:num-lc}
While the bifurcation analyses in Sec.~\ref{sec:analytical-result} is comprehensive 
in regard to the local bifurcations, some global bifurcations such as saddle--node 
bifurcation of periodic orbits are not considered there. These global bifurcations might 
create or destroy stable limit cycles, affecting the stability region of 
the synchronously oscillating state in Fig.~\ref{fig:3-macro-states}.

\begin{figure}[]
   \centering
\includegraphics[]{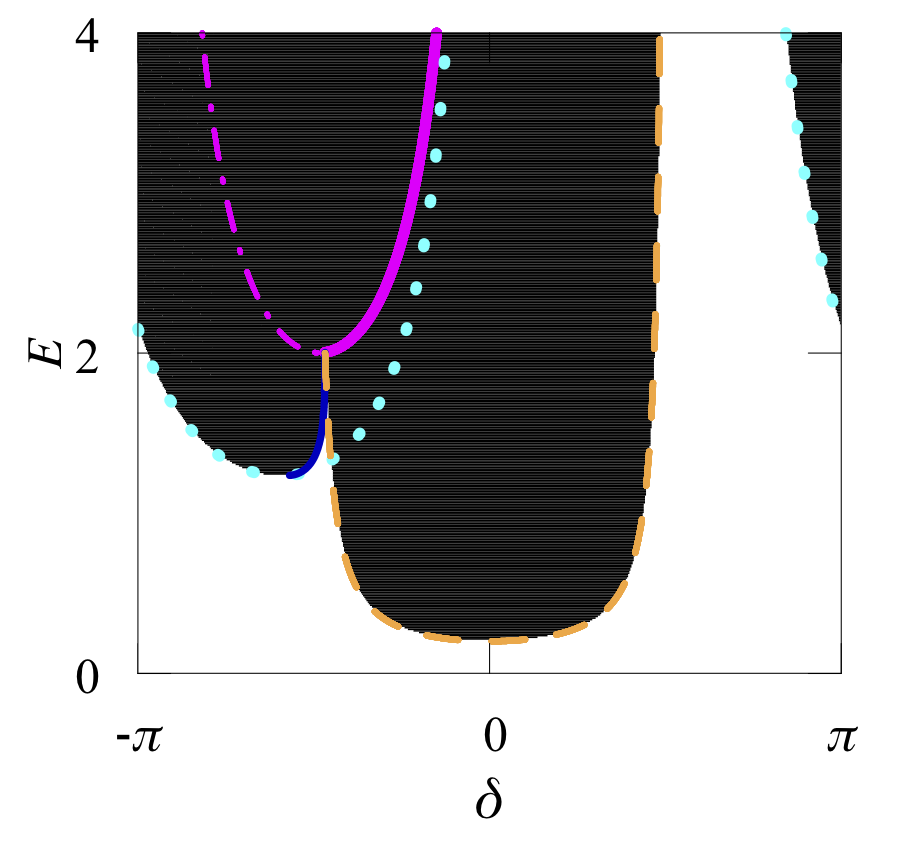}
   \caption{(Color online) The stable region of the limit cycle solution. 
   The region where at least one stable limit cycle exists is filled with black. 
   The bifurcation curves obtained in Sec.~\ref{sec:analytical-result} are also plotted.
   The Hopf, SNIC, and heteroclinic bifurcation curves 
   agree with the boundary of the black region.
   }
   \label{fig:detectlc}
\end{figure}

Thus, we numerically verified that the stability boundary of the synchronously oscillating state consists of 
the Hopf, SNIC, and heteroclinic bifurcation curves obtained in
Sec.~\ref{sec:analytical-result} as detailed below.
For each parameter set, $100$ initial conditions for $\re{r}$ and $\im{r}$ are drawn from uniform distribution on 
the unit disk, and the type of the attractor to which each orbit converges is 
detected. In the black region in Fig.~\ref{fig:detectlc}, at least one orbit converges to
a limit cycle. The edge of the black region agrees with the bifurcation
curves that 
are inferred to form the stability boundary of the collective oscillation.

The type of the attractor is classified according to the following criteria.
(i) Let $\Delta r_t$ be the norm of the variation of the orbit between time $t$ and $t + \Delta t$,
where $\Delta t = 0.01$. The orbit is considered to be converged to a fixed point if 
the phase point is moving slowly ($ \Delta r_{t},~ 
\Delta r_{t-\Delta t},~ \Delta r_{t-2\Delta t}<10^{-5}$) 
and is slowing exponentially
($\left|\Delta r_{t}/ \Delta r_{t-\Delta t}-
 \Delta r_{t-\Delta t} / \Delta r_{t-2\Delta t}\right| < 10^{-5}$).
(ii) Let $x_\mathrm{max}$ and $x_\mathrm{min}$ 
be the maximum and the minimum, respectively, of $\re{r}$ during an interval ($10000<t<20000$). 
Then, let $y_n$ be the $n$th intersection of the line 
$\re{r}=\left(x_\mathrm{max}+x_\mathrm{min}\right)/2$ and the orbit that 
transverses the line with $\re{\dot{r}}>0$. 
The orbit is considered to be converged to a limit cycle when 
$\left|y_n - y_{n-1}\right|< 10^{-5}$.
With this criteria,  
we conclude that every orbit is converged to either  
a fixed point or a limit cycle.

\end{document}